# Exceptional ballistic transport in epitaxial graphene nanoribbons

*Jens Baringhaus[1], Ming Ruan[2], Frederik Edler[1], Antonio Tejeda[3,4], Muriel Sicot[3], AminaTaleb‐Ibrahimi[4], Zhigang Jiang[2], Edward Conrad[2], Claire Berger[2,5], Christoph Tegenkamp[1], Walt A. de Heer[2]\**

[1] Institut für Festkörperphysik, Leibniz Universität, Hannover, Appelstrasse 2, 30167 Hannover, Germany
[2] School of Physics, Georgia Institute of Technology, GA 30332-0430 Atlanta, USA
[3] Institut Jean Lamour, CNRS—Univ. de Nancy—UPV-Metz, 54506 Vandoeuvre les Nancy, France
[4] UR1 CNRS/Synchrotron SOLEIL, Saint‐Aubin, 91192 Gif sur Yvette, France
[5] CNRS-Institut Néel, 38042 Cedex 6, Grenoble, France
in both cases is
* Corresponding author
Jens Baringhaus and Ming Ruan have contributed  ually to this work.

**Graphene electronics has motivated much of graphene science for the past decade. A primary goal was to develop high mobility semiconducting graphene with a band gap that is large enough for high performance applications. Graphene ribbons were thought to be semiconductors with these properties, however efforts to produce ribbons with useful bandgaps and high mobility has had limited success. We show here that high quality epitaxial graphene nanoribbons 40 nm in width, with annealed edges, grown on sidewall SiC are not semiconductors, but single channel room temperature ballistic conductors for lengths up to at least 16 μm. Mobilities exceeding one million corresponding to a sheet resistance below 1 Ω have been observed, thereby surpassing two dimensional graphene by 3 orders of magnitude and theoretical predictions for perfect graphene by more than a factor of 10. The graphene ribbons behave as electronic waveguides or quantum dots. We show that transport in these ribbons is dominated by two components of the ground state transverse waveguide mode, one that is ballistic and temperature independent, and a second thermally activated component that appears to be ballistic at room temperature and insulating at cryogenic temperatures. At room temperature the resistance of both components abruptly increases with increasing length, one at a length of 160 nm and the other at 16 μm. These properties appear to be related to the lowest energy quantum states in the charge neutral ribbons. Since epitaxial graphene nanoribbons are readily produced by the thousands, their room temperature ballistic transport properties can be used in advanced nanoelectronics as well.**

For the past decade, epitaxial graphene bandstructure measurements have been used to demonstrate the "Dirac cone", because its structure and properties are closer to ideal graphene than graphene on any other substrate. The well-known exceptional transport properties of epitaxial graphene (including very high mobilities) and single crystal sizes exceeding hundreds of microns[1] (up to wafer scale[2]) make it an ideal platform for advanced graphene based electronics and scientific investigations of transport near the Dirac point [3,4].

While most graphene research focuses on two-dimensional graphene sheets, here we demonstrate the remarkable properties of graphene nanoribbons and other structures. The ribbons are patterned using high-temperature annealing methods [5] that do not damage the graphene, so that the ballistic transport properties in essentially perfect nanostructures can be observed. Electronic transport in the nanoribbons is strikingly similar to high-quality ballistic carbon nanotubes [6,7]. The experiments and their interpretation are straightforward and provide a glimpse into the fascinating properties of quantum mechanical transport. For example simply touching a ballistic graphene ribbon with a nanoscopic probe reversibly doubles its resistance; touching it with two probes triples it. These expected ballistic transport properties are demonstrated here for the first time, and even more strikingly, they persist at room temperature.

The discovery of room temperature ballistic conduction in epitaxial graphene nanoribbons is important not only for science but also for technology. It allows interconnected ballistic graphene structures to be patterned using conventional microelectronics methods. Further developments, already underway, allow the ballistic currents to be switched with high efficiency, paving the way to integrated epitaxial graphene ballistic nanoelectronics.

These properties are closely related to the electronic structure of the nanoribbons that have widths of about W= 40 nm and lengths on the order L=1 μm. In this size range, they exhibit both waveguide and quantum dot properties. Like photons, a graphene electron moves at a constant speed c*=$10^6$ m/s with momentum p= ℏk, where ℏ is Planck's constant and k is its wave vector. Its energy is E=ℏp so that the energy levels in a W by L ribbon are approximately [8]

$$E_{n,m} = \pm \hbar c^* \sqrt{\left(\frac{n\pi}{W}\right)^2 + \left(\frac{m\pi}{L}\right)^2} \qquad (1)$$

where n and m are integers. For reference if W=40 nm and L=1 μm then $E_{1,0}/k_B$=600 K, $E_{0,1}/k_B$=23 K, where $k_B$ is Boltzmann's constant. Consequently, for temperatures T<600 K only the states with n=0 are occupied even in moderately charged ribbons. Because of the large spacing between quantum states, the ribbons are actually one-dimensional quantum dots (especially at low temperatures) and transport involves the coupling of discrete quantum dot states to the leads. The Dirac point corresponds to the n=0, m=0 state, and may relate to the ballistic transport channel while the activated transport energy is consistent with $E_{01}$ and may relate to the activated channel. The transport properties in quasi-neutral graphene ribbons in the n=0 mode are remarkably distinct from transport in charged graphene in the n≠0 modes.

1. **Side wall nanoribbon production and characterization**

In analogy to high quality carbon nanotubes[6,7], low defect graphene ribbons were expected to be ballistic conductors and ideal interconnects in nanographene circuits [9]. However, disorder [10] in lithographically patterned exfoliated graphene ribbons causes large resistivities and transport gaps on the order of $\Delta E=\pi\hbar c^*/W$ [11-14] (see also Figs. S2) corresponding to the n=0 mode (Eq.1). Well-aligned, single-crystal monolayer graphene sheets form spontaneously on silicon carbide surfaces when heated above 1000°C (as first observed in 1975[15]). Disorder caused by nanolithographic processes is avoided using the structured growth method [5]. In this method graphene ribbons self-assemble on sidewalls of steps that are etched into the (0001) surface of electronics-grade SiC silicon carbide wafers as explained in Refs. [5,16-20] (see Figs 1a and 2).

Single channel ballistic transport was first reported on 40 nm wide graphene ribbons connected to large graphene pads [18,20] (Fig. 1c, see also SI). The graphene ribbon is supplied with a top gate (20 nm $Al_2O_3$ coated with aluminum) and 4 wires are bonded to the graphene leads that seamlessly connect to the nanoribbon, so that 4-point cryogenic (4K≤T≤300K) transport measurements can be performed. Subsequently, in-situ variable geometry transport measurements (at temperatures T from 30-300 K) were performed on ~40 nm wide ribbons (Fig. 1b) confirming single-channel ballistic transport.

The sidewall ribbons have been extensively characterized[5,16-18,20,19] (Figs. 1, 2). The band structure derived from angle-resolved photoemission spectroscopy (ARPES) [19] [21] shows a Dirac cone (Fig. S1), showing that the ribbons are well-aligned monolayers and that the sidewall slope is uniform (i.e. 28°, consistent with the (2-207) facet). The band structure corresponds to charge neutral graphene[19]. Scanning tunneling microscopy (STM) shows the characteristic honeycomb atomic structure of graphene on the sidewalls, and STS shows the characteristic graphene density of states, with a minimum at 0 bias voltage, confirming that these sidewall ribbons are charge neutral (Fig. 2 green framed images). In contrast epitaxial graphene grown on the polar faces is charged. The top plateau shows the semiconducting properties of the "buffer layer" (Fig. 2 red frames). Atomic resolution images edges shows zigzag and chiral edges, consistent with the [1-100] crystallographic alignment of the trenches. Besides their structural integrity, the exact morphology of the graphene ribbon is not essential for our purposes here where we concentrate on transport properties.

**2. In-situ ballistic transport measurements**

An array of 20 nm deep parallel trenches was etched along the [1-100] direction (Fig. 2). After graphitization, graphene decorates the 28° sloped sidewalls to produce 40 nm wide ribbons. While monitored with a scanning electron microscope (SEM) in ultra-high vacuum, up to four nanoscopically sharp tungsten probes are brought into Ohmic (emphatically not tunneling) electrical contact with a selected ribbon (Fig. 1b) and resistances of the individual graphene ribbons are measured. For 2-probe measurements, the current $I_{12}$ is measured between two probes in contact with the ribbon and to which a voltage $V_{12}$ is applied: $R_{2p}=V_{12}/I_{12}$. For 4-probe measurements a current $I_{34}$ is passed through outer probes and the potential difference $V_{12}$ between the two inner probes is measured. The 4-probe resistance of the ribbon segment between the inner probes is $R_{4p}(L)=V_{12}/I_{34}$. For diffusive wires 4-probe measurements are insensitive to the contact resistances, but this is not the case for ballistic conductors. Comparing 2-probe and 4-probe measurements on the same ribbon using the same voltage probes (1) and (2) determines the resistivity ratio RR = $R_{4p}/R_{2p}$ = 0.95±0.02 (see Fig. 3a, middle inset). Furthermore, $R_{4p}-R_{2p}\approx$ 950 Ω = 0.04 $R_0$, where $R_0=h/e^2$=25.8 kΩ. The difference between a 4-probe and 2-probe measurement is typically less than 10% of the measured resistance.

Room temperature 4-probe resistances are found to depend linearly on the distance L between voltage probes for 1μm≤L≤16 μm (Fig. 3a):

R(L)= ΔR+R'L= $R_0(\beta+L/\lambda_{0+})$                           (2a)

where R' is the resistance per unit length, $\lambda_{0+}$ = $R_0$/R'. The R' values range from 0 to 6 kΩ/μm (corresponding to $\lambda_{0+}$>4 μm) and <β>=0.98±0.02. After in-situ annealing (see Fig. 3 and Experimental methods) R' reduces significantly but β≈1 is unchanged.

As shown next, the offset resistance ΔR=β$R_0$ is the quantum contact resistance. For a homogeneous diffusive wire measured in a 4 probe configuration $R_{4p}(L)$=R'L with no offset ΔR. The large offset resistance (ΔR ≈26 kΩ) cannot be explained in terms of diffusive transport, but is expected and well-understood in ballistic conductors as best described in the Landauer formulism (see for instance [22,23]). The Landauer picture provides a convenient connectional framework for transport in quantum wires by resolving the contributions of the modes, (also known as one-dimensional subbands) [22,24]. In graphene ribbons in general, two subbands cross the Fermi level $E_F$ (measured from the graphene charge neutrality point) when $E_F\approx$ 0 [25]. These dominate the conductivity of neutral graphene ribbons (See Fig 4c; for a general discussion of graphene ribbons band structure see Refs.[25-27]; the degeneracy of these two subbands is expected to be lifted by many body effects [28].) Independent of the degeneracy lifting mechanism, the non-degenerate subbands can be labeled by 0+ and 0-. Following Landauer, the conductance G, measured in a 2-probe measurement is

G = $G_0$ ($\mathcal{T}'_{0+}$ + $\mathcal{T}'_{0-}$)                           (2b)

where $G_0=1/R_0=e^2/h$, and $\mathcal{T}'_{0+}$ and $\mathcal{T}'_{0-}$ are the transmission coefficients of these two channels *including* the contacts. If the mean free path is $\lambda_{0\pm}$ then,

$$\mathcal{T}'_{0\pm} = (1+L/\lambda_{0\pm})^{-1} \tag{2c}$$

(ignoring coherence effects [23]). If $L \gg \lambda_{0-}$ then $R(L)= R_0(1+L/\lambda_{0+})$ as observed experimentally. This explains the quantum contact resistance. (See also below.)

The resistance of these nanoribbons (measured for L=5 μm) increases by less than 2% from 30 K to 300 K (as shown in Fig. 3d) and the bias voltage $V_b$ dependence (at T=30 K for L=1.5 μm) is also small (Fig. 3c). It increases by less than 1% from -50 mV to +50 mV, which is consistent with ballistic transport.

In ballistic wires the equivalence of the 4-probe and 2-probe resistance values indicates that the probes are invasive (see, for example, Ref. [24,29]) as demonstrated next.

The invasiveness of the probe is measured by the probability P at which a charge carrier in the ribbon will enter the probe [23,24,29]. For a perfectly invasive probe, P=1 (see also Fig. S9). An invasive probe in contact with a ballistic wire divides it into two ballistic wires, one to the left and one to the right of the probe, each with $R_C = R_0$. Consequently, the end-to-end resistance, $R_{34}$, increases from $R_0$ to 2 $R_0$. Two invasive probes increases $R_{34}$ to $3R_0$. Therefore, for P=1 the resistance ratio RR=$R_{4p}/R_{2p}$ is RR=1 [23,24,29]. We measure RR=0.95 (see above) so that P=0.97 (from P=2RR/(RR+1) [29]). This shows, that our probes are nearly perfectly invasive. (In general, $R_{34}=2R_0/(2-P)$ when a single probe (1) is applied between (3) and (4), and $R_{34}= R_0(P+2)/(2-P)$ when the 2 probes (1) and (2) are applied [23,24,29], see Fig. S9.)

The doubling and tripling effect is clearly observed, as shown in Fig. 3b. The effect is reversible: $R_{34}$ reverts to $R_0$ when the probes are removed. This property is unique to ballistic conductors, since probes (invasive or otherwise) have a negligible affect on the resistance of diffusive conductors. This is the first time that this spectacular demonstration of ballistic conduction has been observed at room temperature and the first time it has been demonstrated with physical probes.

For 0.1 μm≤L≤1 μm, the R(L) is found to increase non-linearly from 0.5 $R_0$ to 1 $R_0$ as shown for 2 ribbons (Fig. 3a, upper left inset). For L< $L^*_{0-}$ = 160 nm R≈$R_0/2$, while for L> $L^*_{0-}$, it decreases (quasi) exponentially, approximately given by

$$G(L) = G_0 \left(1 + \exp-\left(\frac{L}{L^*_{0-}} - 1\right)\right) \tag{3a}$$

(see Fig. 3a upper left inset, and Figs. S7). For the two ribbons R(L) was measured at room temperature from 1 μm to L=25 μm (probe distance limited by the apparatus) (Fig. 3a upper right inset) [30]. A second non-linear increase is observed for L> $L^*_{0+}$=16 μm, and G(L)=$G_0$exp-(L/ $L^*_{0+}$-1) (dashed line in upper right inset of Fig. 3a). Combining the two gives

$$G(L) = G_0\left(\exp-\left(\frac{L}{L^*_{0+}}-1\right)+\exp-\left(\frac{L}{L^*_{0-}}-1\right)\right) \quad (3b)$$

as shown in Fig. 3a (bottom inset). We return to this effect below.

## 3. Transport properties.

Ballistic transport is typically only seen at cryogenic temperatures for well-understood reasons. Elastic scattering can be weak in good samples, but inelastic scattering (i.e. electron-phonon and electron-electron) will become increasingly important with increasing temperature. Micron scale, room temperature ballistic transport is unexplained in current transport theory. In fact our measured room temperature sheet-resistances ($R_{sq}$=R'W) are more than an order of magnitude below the *theoretical* minimum for ideal, freestanding graphene[31] (See Fig. 6). Moreover, they are lower than exfoliated graphene ribbons by up to 4 orders of magnitude and they are below the best 2D freestanding, or BN supported graphene [30] at room temperature by more than 2 orders of magnitude.

Transport measurements are reported on 4 graphene ribbon samples (samples A-D) and on a sidewall graphene ring. Below we show in detail the results for Sample A, Fig. 1c (see SI for B-D). Sample A is a top gated 39 nm wide 1.6 μm long graphene sidewall ribbon. The ribbon is seamlessly connected to micron scale graphene pads to the left and right. Each pad is bonded to two wires facilitating 4-point transport measurements. Resistances in the 20 kΩ range are measured with better than 0.1 Ohm accuracy (corresponding to δG<5 $10^{-6}$ $G_0$). Temperatures are measured with 2 mK accuracy.

As usual, the top gate potential $V_g$ allows the charge density of the ribbon to be varied. (The charge density n($V_g$)=-0.95 $10^{12}$ $V_g$ $cm^{-2}$ $V^{-1}$, as determined from a Hall bar on the same substrate[20], so that $E_F$=-0.11 $V_g^{½}$ $eV$ $V^{-½}$.) As expected, the minimum conductance (G=0.92 $G_0$) occurs near $V_g$=0 for n=0, indicating that the ribbon is not charged by the substrate (consistent with the STS measurements, Fig. 2). The increase in the conductivity with increasing $V_g$ is consistent with the opening of one-dimensional subbands (Fig. 4b) resulting in a conductance increase as always seen in gated graphene ribbons [12,32-34], (Fig. S2). (Occasionally the staircase structure is observed in quantum wires [24], and in graphene ribbons [32] [35].) The conductance increases uniformly with increasing temperature (Fig. 4b) as usually observed in gated graphene ribbons (see Fig S2). It is readily explained in terms of the Landauer picture. At $V_g$=0, the conductance is exclusively due to the n=0 subbands. Increasing the gate voltage increases the Fermi level in the ribbon. When it rises above the n=1 subband minimum, $E_1$, the conductance increases by $4T_1G_0$, where $T_1 \leq 1$ is the transmission coefficient (see Fig. 4c). Subsequent increases successively occur for larger n (as usual the steps are not well resolved). Since the conductance increases uniformly with increasing temperature implies that the n=0 subband dominates this property (as pictorially demonstrated in Fig. 4d).

The weak temperature dependence on the n≠0 subbands corresponds to well-known properties of charged 2D graphene. The large asymmetry with respect to $V_g$ (Fig. 4a) is caused by the np/pn junctions (see for example [36]). For $V_g$ <0 the ribbon is p doped while the leads are slightly n doped [20]. For our ribbons $\lambda_{|n|>0} \approx 50$ nm (as determined from the $dG/dV_g$) is larger than the junction width (that is on the order of the dielectric thickness, i.e. 20 nm). Consequently, in these top gated ribbons, the junctions represent a significant barrier for the n≠0 subbands[36] (Fig. 4 c, d). As for exfoliated ribbons the n≠0 bands are diffusive on the 100 nm scale with a small temperature dependence, (like non-neutral two-dimensional graphene). Note that in exfoliated graphene ribbons $\lambda_{|n|>0} \approx 5$ nm have a mobility gap is [10,12,14,32-34] (see SI; Fig S2). The mobility gap is due to the very large disorder that dominates transport in the n=0 subband [12,13] (Figs. S1,S2). In contrast, epitaxial graphene ribbons are atomically flat [37], they have little structural disorder [19] (Fig. S1), no trapped charge puddles[37] (potential variations within a few meV over 200 nm) and no mobility gap (Fig. S2). That is why they are suited for investigations of Dirac point physics [3,4].

From the above discussion it is clear that quantum confinement in the ribbon opens an energy window $\Delta E = 2E_1$ about $E_F=0$ where only the n=0 subbands contribute to the transport (see Fig. 4c,d) The remainder of this paper focuses exclusively on them.

The conductance G(T) of Sample A was continuously monitored while it was slowly cooled from T=120 K to 4.3 K. G(T) monotonically decreased from 1.22 $G_0$ to 0.94 $G_0$. Analysis shows that the conductance is very well described by:

$$G(T) = \alpha \frac{e^2}{\hbar} \left( 1 + \frac{1}{2} \exp - \left( \frac{T^*}{T_{el} - T_0} \right)^{1/2} \right)$$

$$T^* = 4\hbar c^* / k_B L$$

$$T_{el} = \sqrt{T^2 + \left(eV_b/vk_B\right)^2}$$

$$T_{el} = T + \left|\gamma \mu_B B / k_B\right|$$

(4a-d)

where $c^*=1.0\ 10^6$ m/s is the Fermi velocity; $V_b$ is the bias voltage, B is the magnetic field (the magnetic field and bias voltage dependence is discussed below). For Sample A, $\alpha$=0.922, $T^*$=21.13 K and $T_0$=2.2 K. From $\alpha$ and Eq. 2 we determine that $\lambda_0$ =21 μm (i.e. clearly ballistic). The difference $\delta G(T)$ between the fit and data are shown in Fig. 5a (lower inset). As shown in Fig. 5a, the fit is accurate to within 0.0015 $G_0$ (0.1%) over the entire range from T=4 K to T=120K (Fig. 5a).

Equation 4 applies to Samples A-D (data for B-D is less extensive, fits involve $\alpha$ and T*, using the experimental L).
(1) For sample A from G(T) (Fig. 4a), the measured $T_m^*$=21.3 K (Fig. 4a). From Eq. 1b, with L=1.6 µm, the predicted $T_p^*$=19.0, $\alpha$=0.92
(2) For sample B (Fig. S4) $\alpha$=0.31; $T_m^*$=29 K ; for L=1.06 µm, $T_p^*$=28 .
(3) For sample D (Fig. S5), $\alpha$=0.63; $T_m^*$=88 K; fot L=0.36 µm, $T_p^*$=84.5 K .
(4) For the ring structure (Fig. 1d), $T_m^*$=6±1 K, the measured contact-to-contact distance is 5 µm (following ½ a turn of the circle), from which $T_p^*$=6.1 K.
This clearly establishes the inverse L dependence of T* in these samples ($\alpha$ and T* appear to be unrelated).

The conductance increases with increasing bias voltage $V_b$ (Fig. 5b). This effect has been attributed to electronic heating [12]. Since the relationship between temperature and conductance is accurately known (Eq. 4a-c), the conductance can be used as a thermometer for the electronic temperature $T_{el}$ (that typically differs from the lattice temperature T at non-zero bias voltage). The conductance $G(V_b)$ is plotted in terms of $T_{el}$, by converting $G(T_{el})$ to $T_{el}$ (G), and the $V_b$ axis is scaled to its equivalent temperature $T_{Vb}=eV_b/k_B$ (Fig. 5b), thereby representing the energy of each electron injected from the contact, by its equivalent temperature. When plotted this way, the data accurately describe hyperbolae (Eq.4c):

$$T_{el} = \sqrt{T^2 + \left(eV_b/\nu k_B\right)^2}$$

where $\nu$ is dimensionless. Figure 5c shows a good fit for all temperatures with $\nu = 5$ for $T_{el}$< 15 K and $\nu_2$ =12 for $T_{el}$> 15 K. This behavior is observed in all of our samples and in a carbon nanotube. [38]. (See SI and Fig. S8)

Magnetotransport properties are described next (Fig. 5d). Following the method above, the magnetoconductance G(B) is converted into its corresponding temperature $T_{el}$(G) and B is converted to $T_B=\mu_B B/k_B$ (Fig. 5e). When plotted this way, the (dimensionless) slopes µ= $dT_{el}/dT_B$ represent effective magnetic moments in units of the Bohr magneton $\mu_B$. Note the essentially perfectly linear, V-shaped, about 4T wide, magnetoconductance dip. The dip corresponds to $T_{el}$=T+µ|B| (Eq.4d) where µ = 5$\mu_B$ at low temperatures and increases to µ ≈ 10 $\mu_B$ at high temperatures. For all ribbon samples at all temperatures the constant slope abruptly changes at |B|≈ 2T (Figs. S3-6). For the graphene ring (Fig. 1d) the magnetoresistance is closely related (Fig. 5f,g). Figure 5g shows that $T_{el}$ versus $T_B$ consists of linear segments. For the vertical dip, µ = 2.13 $\mu_B$. For the other two segments, µ = 0.869 $\mu_B$ and µ = 0.827 $\mu_B$.

These properties are also seen in high-quality carbon nanotubes [38]. Conductance measurements as a function of T, B and $V_b$ were made on a ≈ 10 nm diameter nanotube contacted with 4 lithographically patterned probes (Schönenberger et al.[38] ). The center-to-center probe distance was L = 350 nm. The conductance as a function of temperature follows a modified version of Eq. 4 ($e^2/h$ is replaced by $2e^2/h$ and the prefactor ½ is replaced by 3) with α=0.84, T*= 65 K and $T_0$ ≈ 1 K (see Fig. S8). The conductivity for $V_b$<5mV is described by Eq. 4c with ν≈4. For B<2T the conductance is described by Eq. 4d with γ≈ 10. Moreover, room temperature micron scale ballistic conduction with $G_{0+}=2e^2/h$ has been observed in high quality carbon nanotubes [6]. The increase by a factor of 2 of the conductance compared with graphene ribbons is consistent with charge neutral carbon nanotubes.

## 4. How are charge carriers in neutral graphene ribbons different?

Ballistic transport is only seen in the n=0 subband with room temperature conductivities exceeding the reported exfoliated graphene measurements by up to 2 orders of magnitude. It even exceeds predictions of ideal graphene by a more than a factor of 10 [31,39] (Fig. 6). There are two distinct charge carriers in the n=0 subband. One is a single channel room micron scale ballistic conductor that is bias voltage and temperature independent. The other is temperature, magnetic field and bias voltage dependent. Both exhibit similar non-linear conductance increases (at room temperature), one at 160 nm and the other at 16 μm. In contrast in the same ribbon, the mean free paths in the n≠0 subbands, is on the order of 50 nm. The n=0 subband is clearly special. If it very important to realize that the ballistic properties reported here are not unique to epitaxial graphene but that they are also seen in high quality carbon nanotubes. This, combined with the fact that ballistic transport is seen in meandering ribbons, implies that ballistic transport is not associated with specific (i.e. zigzag) edge morphologies.

The transport properties reported in Fig. 3 agree with those reported in Fig. 4, both showing two-component behavior. In both cases the relevant energy scale is hc*/L. However there are significant differences in the experimental conditions. In Fig. 3 invasive contacts were placed directly on the ribbon, so that the voltage probes are ultimately the source and drain for the particles that pass through the ribbons, thereby establishing a well-defined fixed path between the two. On the other hand, for samples in Fig. 4 the graphene ribbon is seamlessly connected to extended wide (low resistance) graphene pads that ultimately connected to metal leads much further away, thereby allowing variability in the conductance path.

Since Eq. 4 resembles the expression for one-dimensional variable range hopping, it is reasonable to explore the possibility of a related mechanism. Following Mott's heuristic argument[40], charge carriers select the path of least resistance, i.e. one with the largest transmission coefficient. In our case, the path passes through the ribbon for a total distance of R. The transmission coefficient involves the product Z of two competing terms. One is the Boltzmann factor to occupy the lowest longitudinal mode $E_0=\pi\hbar c^*/R$ and the other is the determined by the lifetime of the charge carrier $\tau$. Consequently $Z=\exp(-E_0/k_BT)\exp(-t/\tau)=\exp(-\pi\hbar c^*/Rk_BT)\exp(-R/c^*\tau)$. Note that Z maximizes for $R=\sqrt{(\pi\hbar c^{*2}\tau/k_BT)}$ so that $Z=\exp-\sqrt{(T^*/T)}$, where $T^*= \pi\hbar/k_B\tau$. From Eq. 4, we conclude that $\tau=\pi L/c^*$, (i.e. on the order of the transit time through the ribbon). This indicates that transport involves thermally activated longitudinal modes of the graphene ribbon. Consequently T* is not a material property but depends inversely on the ribbon length. The mechanism above implies coherence lengths comparable to L (up to at least T=150 K, Fig.5a).

The high quality of epitaxial graphene on SiC, compared with exfoliated and transferred graphene is well known [1] and that is why epitaxial graphene is used to demonstrate prototypical graphene properties [3,41]. The already small residual substrate interaction on the polar faces (whose main, if not only effect is to induce a uniform charge density on the order of $10^{12}/cm^2$) is further reduced in sidewall graphene, where the graphene is charge neutral and slightly separated from the substrate[42]. This, combined with the annealed edge structure, produces the nearly ideal graphene ribbons reported here, with transport properties exceeding other forms of graphene (Fig. 6).

In contrast, in transferred and suspended graphene, charge inhomogeneity and contamination (interfacial charge puddles and trapped nanoparticles) is unavoidable (Fig. S1). Uncontrolled morphology (amorphous substrates, uncontrolled substrate adhesion, disordered edges and random strain [43]) profoundly affect transport especially at low charge densities, causing a mobility gap in the n=0 subband and other spurious effects (Figs. S1, S2). This affects even large samples as is clear from Eq. 1. For example if the edges of an of a charge neutral square graphene sample with L=2 μm are disordered, then a mobility gap is be expected below 12 K, consistent with experiment [44].

**Summary and Conclusion**

Room temperature single channel ballistic conduction in the n=0+ subband was first observed 2 years ago in epitaxial graphene sidewall ribbons[20]. Although experimental evidence is overwhelming, conventional transport theory has not found a generally accepted explanation, neither in graphene ribbons nor in nanotubes. In summary single channel ballistic transport in the n=0+ channel is supported in

(1) the robust value of the quantum contact resistance;
(2) the length independence;
(3) the temperature independence;
(4) the bias voltage independence;
(5) the resistance doubling and tripling due to passive probes, and
(6) the equivalence of 2 and 4 probe conductance measurements.

We demonstrated ballistic transport in more than 50 ribbons (SI Fig.S12 and Table 1). The ribbons are shown to be charge neutral monolayers with well-formed edges. Its absence in exfoliated graphene is due to the high degree of disorder in that material.

The detailed analysis of the n=0- channel shows that it is more complex. The dependence of the conductance on temperature is clearly caused by thermal activation and not by disorder. The activation barrier is related to the longitudinal modes in the ribbon. The dependence on bias voltage indicates that impinging electrons produce hot charge carriers that overcome the activation barrier. Analysis shows that the n=0- charge carriers are associated with magnetic moments on the order of 5 $\mu_B$. It is interesting to note that the effective magnetic moment of the 1/3 fractional quantum Hall state[45], $\mu=d\Delta\mu_v/dB$ corresponds to $\mu=5.7$ $\mu_B$. Moreover, Miller et al. [37]observed a linear magnetic field dependence in the 0 Landau level in epitaxial graphene corresponding to $\mu=10$ $\mu_B$.

The magnetic field dependence and bias voltage dependence of the n=0- subband are well described by the parameters µ and ν that are consistent from one ribbon to the next. The parameter α probably quantifies elastic scattering. It affects both n=0± subband channels similarly and it is closer to unity in ultrahigh vacuum and for graphene ribbons protected by a dielectric, than for ribbons exposed to ambient conditions. In exfoliated graphene ribbons the mobility gap in the n=0 subband is reflected in the very small value of α≈$10^{-2}$ as discussed above (see also Figs. S1, S2).

An intriguing property of the n=0± channels is the abrupt non-linear conductance increase at $L_{0-}$=160 nm for n=0- and $L_{0+}$=16,000 nm for n=0+ at room temperature. At T=300 K the electron wavelength $\Lambda_{th}$= hc*/$k_B$T=159 nm, suggesting that at these temperatures, the resistance increase is related to the longitudinal modes. However it is difficult to apply this argument to the n=0+ as well. Although it may be that the n=0+ state represents the Dirac point (n=m=0) where velocity renormalization[46] [47] may be extreme (a factor of 100). It is also possible that both the 0- and 0+ states relate to quasiparticles resulting from the symmetry broken ground state of graphene [47], that decay at different (energy dependent) rates. This intriguing possibility may also explain the $T_0$ term (Eq. 4).

These explanations are actually in line with predicted non-Fermi liquid properties of the graphene ground state that has been extensively debated [47]. However, this unusual state of matter has been inaccessible experimentally due to its demanding requirements on sample purity and structural homogeneity. The experimental evidence that we provided indicates that requisite conditions are met in our epitaxially grown graphene samples with well-annealed edges, and particularly so in the experiments performed in ultrahigh vacuum conditions.

There is no doubt that the ultimate explanation of the exceptional transport properties reported here will significantly affect our understanding of ground state graphene. Moreover graphene structures with these properties are readily produced on commercially available, relatively low cost[1] electronics grade silicon carbide. This opens the door to electronics based on ballistic transport at room temperature; a paradigm changing development may affect the direction of future electronics.

**Experimental methods**

For samples in Fig 2, 4 and 5, graphene ribbons were produced by thermally annealing[16] either natural or etched steps on the (0001) face of electronics grade insulating 4H-SiC[5,17,18,20]. 20 nm deep trenches, aligned along the [1-100] direction, were etched and annealed at 1600°C for 15 min.[20] Samples in Fig. 3 were heated at 1300°C in Ar (4 10$^{-5}$ mb), and then in UHV for 15 min at 1100°C (green dots, Fig. 3b) or at 1150°C (all others – for details see SI, Part B). The natural step samples (like in Fig. 4) were prepared on chips that were provided with two 200 nm deep trenched separated by 1 µm with a natural step connecting the two.

AFM and conducting-AFM (C-AFM) were used to measure the ribbon widths (see[20]) and to measure the sidewall slopes. SEM, AFM, Electrostatic Force Microscopy (EFM) and C-AFM were used for characterization. The Soleil synchrotron facility (Cassiopée ARPES beam line) was used for graphene band structure measurements (Fig. S1) to determine the number of graphene layers and to measure the sidewall slopes[20] .

Low temperature (T=77K) STM-STS measurements (Fig. 2) were performed (in Nancy) in a UHV chamber coupled to a surface science dedicated preparation chamber and a photoemission set-up. Bias voltages indicated for STM data correspond to sample potentials with respect to the grounded tip. STS spectra were acquired with a PtIr tip under a bias modulation of 70 mV at 1100 Hz and a lock-in detection of the tunnel current, in open feedback loop conditions.

An Omicron Nanoprobe UHV system (in Hannover), for temperatures from 30 K to 300 K (using tungsten tips) was used for the multi-probe measurements in Fig. 3. The set point for STS (Fig. 1d) was 2V/0.1nA. Probes were positioned using a built-in SEM (Fig. 3a). The resistance between neighboring ribbons was > 500 kΩ (Fig. S11). A Janis variable temperature cryostat system (for temperatures from 4 K to 300 K) with a 9 T magnet was used for the measurements (in Atlanta) reported in Fig. 4-5. Samples were measured using standard low frequency lockin measurement techniques (13Hz, 100nA<I<1µA) [20]. G(T) measurements were performed by cooling from 120K to 4K over a period of 10 hours, with a measurement rate of 2 measurements/s.

**Acknowledgements**

CT thanks the German Research Foundation Priority Program 1459 'Graphene' for financial support. CB, EC, and WdH thank Rui Dong, Paul Goldbart, Zelei Guo, John Hankinson, Jeremy Hicks, Yike Hu, Jan Kunc, An-Ping Li, Markus Kindermann, Didier Mayou, Meredith Nevius, James Palmer, Anton Sidorov for assistance and comments. CB, EC, and WdH thank the AFOSR, NSF (MRSEC – DMR 0820382), W.M. Keck foundation and Partner University Fund for financial support.

JB and FE produced samples and performed the transport experiments relating to Figs. 3. CT performed and conducted the transport experiments in Fig. 3, discussed the data and commented on the paper. MR produced the samples and performed all transport experiments except those relating to Fig.3. EC, AT and ATI performed ARPES experiments, and AT and MS the STM and STS experiments. ZJ performed confirming spin transport measurements. WdH performed the data analysis. WdH, CB and MR designed the project, performed the Atlanta based experiments and wrote the paper.

References

1. Novoselov, K.S. et al. A roadmap for graphene. *Nature* **490**, 192-200 (2012).
2. de Heer, W.A. et al. Epitaxial graphene. *Solid State Communications* **143**, 92-100 (2007).
3. Geim, A.K. Graphene: Status and Prospects. *Science* **324**, 1530-1534 (2009).
4. Orlita, M. et al. Approaching the Dirac Point in High-Mobility Multilayer Epitaxial Graphene. *Physical Review Letters* **101**, 267601 (2008).
5. Sprinkle, M. et al. Scalable templated growth of graphene nanoribbons on SiC. *Nature Nanotechnology* **5**, 727-731 (2010).
6. Frank, S., Poncharal, P., Wang, Z.L. & de Heer, W.A. Carbon nanotube quantum resistors. *Science* **280**, 1744 (1998).
7. Purewal, M.S. et al. Scaling of resistance and electron mean free path of single-walled carbon nanotubes. *Physical Review Letters* **98**, 186808 (2007).
8. Berger, C. et al. Electronic confinement and coherence in patterned epitaxial graphene. *Science* **312**, 1191-1196 (2006).
9. Berger, C. et al. Ultrathin epitaxial graphite: 2D electron gas properties and a route toward graphene-based nanoelectronics. *Journal of Physical Chemistry B* **108**, 19912-19916 (2004).
10. Stampfer, C. et al. Transport in graphene nanostructures. *Frontiers of Physics* **6**, 271-293 (2011).
11. Han, M.Y., Özyilmaz, B., Zhang, Y. & Kim, P. Energy Band-Gap Engineering of Graphene Nanoribbons. *Physical Review Letters* **98**, 206805 (2007).
12. Han, M.Y., Brant, J.C. & Kim, P. Electron Transport in Disordered Graphene Nanoribbons. *Physical Review Letters* **104**, 056801 (2010).
13. Oostinga, J.B., Sacepe, B., Craciun, M.F. & Morpurgo, A.F. Magnetotransport through graphene nanoribbons. *Physical Review B* **81**, 193408 (2010).
14. Chen, Z.H., Lin, Y.M., Rooks, M.J. & Avouris, P. Graphene nano-ribbon electronics. *Physica E-Low-Dimensional Systems & Nanostructures* **40**, 228-232 (2007).
15. Van Bommel, A.J., Crobeen, J.E. & Van Tooren, A. LEED and Auger electron observations of the SiC(0001) surface *Surface Science* **48**, 463 (1975).
16. de Heer, W.A. et al. Large area and structured epitaxial graphene produced by confinement controlled sublimation of silicon carbide. *Proc Nat Acad Sci* **108**, 16900-16905 (2011).
17. Hu, Y.K. et al. Structured epitaxial graphene: growth and properties. *Journal of Physics D-Applied Physics* **45**, 154010 (2012). (Note the magnetoconductance oscillations were only approximately labeled in this paper).
18. Ruan, M. et al. Epitaxial graphene on silicon carbide: Introduction to structured graphene. *MRS Bulletin* **37**, 1138 (2012).
19. Hicks, J. et al. A wide-bandgap metal–semiconductor–metal nanostructure made entirely from graphene. *Nature Physics* **9**, 49-54 (2013).

20. Ruan, M. PhD dissertation- Georgia Institute of Technology - http://hdl.handle.net/1853/45596 (July 2012. Ballistic conduction in epitaxial graphene nanoribbons due to the edge states was first reported in measurements performed at Georgia Institute of Technology. This inspired the extensive and thorough research program reported here, involving three research teams on two continents).
21. Ohta, T. et al. Interlayer interaction and electronic screening in multilayer graphene investigated with angle-resolved photoemission spectroscopy. *Physical Review Letters* **98**, 206802 (2007).
22. Beenakker, C.W.J. & Vanhouten, H. Quantum Transport in Semiconductor Nanostructures. *Solid State Physics-Advances in Research and Applications* **44**, 1-228 (1991).
23. Datta, S. *electronic transport in mesoscopic systems*, (Cambridge University Press, Cambridge, 1995).
24. de Picciotto, R., Stormer, H.L., Pfeiffer, L.N., Baldwin, K.W. & West, K.W. Four-terminal resistance of a ballistic quantum wire. *Nature* **411**, 51-54 (2001).
25. Wakabayashi, K., Takane, Y. & Sigrist, M. Perfectly conducting channel and universality crossover in disordered graphene nanoribbons. *Physical Review Letters* **99**, 036601 (2007).
26. Nakada, K., Fujita, M., Dresselhaus, G. & Dresselhaus, M.S. Edge state in graphene ribbons: Nanometer size effect and edge shape dependence. *Physical Review B* **54**, 17954-17961 (1996).
27. Castro Neto, A.H., Guinea, F., Peres, N.M.R., Novoselov, K.S. & Geim, A.K. The electronic properties of graphene. *Reviews of Modern Physics* **81**, 109-162 (2009).
28. Yazyev, O.V. Emergence of magnetism in graphene materials and nanostructures. *Reports on Progress in Physics* **73**, 056501 (2010).
29. Buttiker, M. Four terminal phase coherent conductance. *Physical Review Letters* **57**, 1761 (1986).
30. Mayorov, A.S. et al. Micrometer-Scale Ballistic Transport in Encapsulated Graphene at Room Temperature. *Nano Letters* **11**, 2396-2399 (2011).
31. Das Sarma, S., Adam, S., Hwang, E.H. & Rossi, E. Electronic transport in two-dimensional graphene. *Review of modern physics* **83**, 407 (2011).
32. Lin, Y.M., Perebeinos, V., Chen, Z.H. & Avouris, P. Electrical observation of subband formation in graphene nanoribbons. *Physical Review B* **78**, 161409(R) (2008).
33. Wang, X.R. et al. Graphene nanoribbons with smooth edges behave as quantum wires. *Nature Nanotechnology* **6**, 563-567 (2011).
34. Todd, K., Chou, H.T., Amasha, S. & Goldhaber-Gordon, D. Quantum Dot Behavior in Graphene Nanoconstrictions. *Nano Letters* **9**, 416-421 (2009).
35. Tombros, N. et al. Quantized conductance of a suspended graphene nanoconstriction. *Nature Physics* **7**, 697-700 (2011).
36. Huard, B. et al. Transport measurements across a tunable potential barrier in graphene. *Physical Review Letters* **98**, 236803 (2007).
37. Miller, D.L. et al. Observing the Quantization of Zero Mass Carriers in Graphene. *Science* **324**, 924-927 (2009).

38. Schonenberger, C., Bachtold, A., Strunk, C., Salvetat, J.P. & Forro, L. Interference and Interaction in multi-wall carbon nanotubes. *Applied Physics a-Materials Science & Processing* **69**, 283-295 (1999).
39. Chen, J.H., Jang, C., Xiao, S.D., Ishigami, M. & Fuhrer, M.S. Intrinsic and extrinsic performance limits of graphene devices on SiO2. *Nature Nanotechnology* **3**, 206-209 (2008).
40. Mott, N.F. Conduction in non-crystalline materials. III. Localized states in a pseudogap and near extremities of conduction and valence bands. *Phil. Mag* **19**, 835-52 (1969).
41. Sprinkle, M. et al. First Direct Observation of a Nearly Ideal Graphene Band Structure. *Physical Review Letters* **103**, 226803(4pp) (2009).
42. Nicotra, G. et al. Delaminated Graphene at Silicon Carbide Facets: Atomic Scale Imaging and Spectroscopy. *Acs Nano* **7**, 3045-3052 (2013).
43. Bao, W.Z. et al. Controlled ripple texturing of suspended graphene and ultrathin graphite membranes. *Nature Nanotechnology* **4**, 562-566 (2009).
44. Ponomarenko, L.A. et al. Tunable metal-insulator transition in double-layer graphene heterostructures. *Nature Physics* **7**, 958-961 (2011).
45. Feldman, B.E., Krauss, B., Smet, J.H. & Yacoby, A. Unconventional Sequence of Fractional Quantum Hall States in Suspended Graphene. *Science* **337**, 1196-1199 (2012).
46. Hwang, C. et al. Fermi velocity engineering in graphene by substrate modification. *Scientific Reports* **2**(2012).
47. Kotov, V.N., Uchoa, B., Pereira, V.M., Guinea, F. & Castro-Neto, A.H. Electron-Electron Interactions in Graphene: Current Status and Perspectives. *Reviews of Modern Physics* **84**, 1067-1125 (2012).
48. Norimatsu, W. & Kusunoki, M. Formation process of graphene on SiC (0001). *Physica E-Low-Dimensional Systems & Nanostructures* **42**, 691-694 (2010).
49. Huard, B., Stander, N., Sulpizio, J.A. & Goldhaber-Gordon, D. Evidence of the role of contacts on the observed electron-hole asymmetry in graphene. *Physical Review B* **78**, 121402R (2008).
50. Lemme, M., Echtermeyer, T.J., Baus, M. & Kurz, H. A graphene field effect device. *IEEE Electron Device Letters* **28**, 282–284 (2007).

Figure Captions

Figure 1

Structure and characterization of epitaxial graphene sidewall nanoribbons and devices. (a) Schematic diagram of a graphene ribbon on an annealed and facetted sidewall, showing the seamless connection to the covalently bonded semiconducting buffer layer on the top terrace and the covalent bonded graphene edge at the bottom edge (see [48]) . (b) AFM image of an array sidewall graphene covered with 20 nm deep trenches (bottom inset shows 3D view, vertical dimension is magnified, rounding is due to tip effects). (top inset) Guided by a SEM, up to 4 individual probes are brought into contact with a selected graphene ribbon. The sample can be transferred to and from an in situ heating stage for annealing up to 1500°C. (c) Optical micrograph of sidewall ribbon (Sample A) supplied with leads and gate consisting of wide graphene ribbons (1 μm apart) connected by a nominally 39 nm ribbon to form an "H" shaped geometry, where the vertical, wide graphene ribbons serve as leads to the 1.6 μm long ribbon. White dashed lines indicate location of the graphene leads, white line indicates graphene ribbon. Green region locates the gate structure. Dark areas are the gold contacts. (d) Electrostatic force image of a sidewall graphene nanoring 1.6 μm outer diameter attached to graphene leads. The ring is produced similarly to (c) and has graphene covered sloping sidewalls.

Figure 2

Scanning tunneling analysis of ex-situ produced sidewall ribbons similar to those used in fixed geometry transport measurements. Color coded line over 28º slope indicates areas of the surface investigated, corresponding to the images in the colored frames. (upper middle) Atomic resolution STM shows graphene structure on the sloped sidewall and corresponds to the typical graphene STS (bottom middle). STM of upper (middle left) and lower (middle right) edges show helical edge structures. STM of the upper (upper left) and lower (upper right) terraces. Corresponding STS of those terraces (lower left, lower right) show semiconducting gap.

Figure 3

Multiprobe in-situ transport measurements of 40 nm wide graphene sidewall ribbons. (a) Resistances as a function of voltage probe spacing L. Linear fits extrapolate to $R_0 = h/e^2$, within a few percent at L=0. Slopes correspond to R'= 6.2, 1.6, 0.92, 0.44 (-0.25) kΩ/μm from 1 to 5 corresponding to $\lambda_0$= 4.2, 28, 16, 58, (-100) μm respectively. 1: UHV annealed at 1100°C for 15 min, 2: UHV annealed at 1150°C for 15 min. 3-5: re-annealed at 1150°C for 15 min. (see Experimental Methods) (Middle inset) A resistance increase of (only) about 4% is observed for 2-probe compared with 4-probe measurements indicating almost perfectly invasive contacts (R(L=0)=0.97 $R_0$; $\lambda_0$ =106 μm). (Upper inset) Resistances of 2 ribbons measured at room temperature showing similar non-linear increases for L>16 µm; for small L, R'=46 and 28 Ω/µm (corresponding to 1.8 and 1.1 Ω/square at least a factor of 1000 smaller than neutral 2D graphene), and $\lambda_0$= 560 and 900 µm. (Lower inset) Resistances of 2 ribbons measured at room temperature showing similar $\sim R_0/2$ resistances for L<160 nm and non-linear increases for L>1600 nm, dashed line corresponds to R=$R_0$/(1+(exp(1-L/Λ)) where Λ= 160 nm. The theoretical fit is from Eq.3b. (b) Effect of passive probes invasively contacting ballistic graphene ribbons. A single passive probe essentially doubles the two-point resistance $R_{ab}$ between the a and b ends of the ribbon; two passive probes triples it, showing that the passive probes are invasive, explaining why 4 probe and 2 probe resistances yield essentially identical values. The shaded area and open circles indicate theoretical limits for an ideal invasive probe (P=1) and a non-invasive probe (P=0). The theoretical [24, 29] values (using RR=0.95) are indicated by * .(c) Resistance as a function of bias voltage $V_b$ showing essentially no effect for -50 mV≤$V_b$≤50 mV (d) Resistance as a function of temperature (L=5µm) showing less that 10 % variation from 30 K to 300 K .

Figure 4.

Transport in graphene gated sidewall ribbon. (a) Conductances $G_T(V_g)$ for various temperatures as shown. Minimum conductance at $V_g$=0 corresponds to charge neutral ribbons ($E_F$=0) where only the n=0 energy states contribute; The conductance increases for $V_g$>0 and less so for $V_g$<0; (b) Decomposition of $G_T (V_g)$=$G_{0+}$ + $G_{0-}$(T) + $G_n(V_g)$ (coded in blue, red and green respectively). Only the 0+ and 0- states contribute at Vg=0 (where $E_F$=0). Only the 0- subband shows a temperature dependence. The |n|≥1 bands ($E_F$≠0) show no temperature dependence (apart from weak oscillations), as seen by the collapse of all the curves on a single one. At T=0 and $E_F$=0 only 0+ state contributes. (c) Schematic band structure of a chiral graphene ribbon; n labels the subbands. For |$E_F$|≤$E_1$ only the n=0± subbands contribute to transport. (d) Diagrammatic representation of the Landauer formula. The (gapless) 0± subbands contribute for all $E_F$; the conductance rise for increasing $V_g$ is due to the |n|≥1 subbands; the conductance increase with increasing temperature is due to the 0- subband; the 0+ subband provides a constant contribution (≈$G_0$), independent of $E_F$ and temperature.

Figure 5.

Example of transport properties of a typical epitaxial graphene ribbon (Fig. 1c) and an epitaxial graphene nanoring. (Fig.1d) demonstrating Eq. 4 (with its B and $V_b$ dependences). Data presented are unprocessed. (a) 4 point G versus T measured over a 10 hour time period (red curve), superimposed on the theoretical curve from Eq. 4a (blue curve) showing an excellent fit. (Upper inset), plot of (T-$T_0$)/T* versus T for the data in the main panel. The line intercept T=0 at $T_0$=2.2 K, and its inverse slope is T*=21.5 K. (Lower inset) Difference δG(T) of Eq. 4a and experiment. (b)

Differential conductance $dI/dV_b$ versus bias voltage $V_b$ of Sample A, from bottom to top, T=4.3K, 7K 12K, 20K, 35K, 55K, 80K. (c) Same data as in (b) plotted as a function of $T_{el}$, where the conductance is converted into temperature following the data in (a), and $T_{vb}=eV_b/k_B$. Fits correspond $T_{el}=\sqrt{(T^2+(T_{vb}/\nu)^2)}$. For $T_{el}$<15 K, $\nu$= 5; for larger $T_e$, $\nu$=10. (d) Conductance versus magnetic field B of Sample A. Temperatures are (from bottom to top) 4.3 K, 7K, 12K, 20K, 35K, 55K, 80K. (e) Same data as in (d) plotted as a function $T_{el}$ and $T_B= \mu_B B/k_B$. Slopes correspond to the magnetic moments in units of $\mu_B$. For B<2T, these are 5.6, 6.3, 8.3, 9.7, 11.3, 10.8, 10.4 and 11. (f) Conductance versus magnetic field for graphene nanoring (Fig. 1d), T=4.3 K (two overlapping sweeps with corresponding linear fits are displayed). (g) Same data as in (f) plotted as $T_{el}$ versus $T_B$. Slopes correspond to $\mu$= ±2.13 $\mu_B$, $\mu$ = ±0.87 $\mu_B$, $\mu$= ±0.83 $\mu_B$. The fits intercept B=0 below $T_e$=4.3 K.

Figure 6.

Transport properties of epitaxial graphene nanoribbons reported here (5) and (6) compared with exfoliated two-dimensional graphene: (1) on $SiO_2$ ref.[39], (2) ultrahigh mobility graphene on BN [30], (3) bulk silver, (4) the theoretical ideal graphene limit [39]; and lithographically prepared graphene nanoribbons at a charge density $n_s$ = $10^{12}$cm$^{-2}$ at T=30K (7) Ref. [11], (8) Ref. [14], (9) Ref. [49], (10) Ref. [50], (11) ref. [32], (12) [12], (13) Ref. [34], (14) Ref. [33]. Sheet resistances ($R_{square}$) for graphene ribbons are determined by multiplying reported resistances (in $R_0$ units) by W/L. For back-gated graphene the resistivity at $V_g=V_D±14$ V (corresponding to $n_s$ = $±10^{12}$ cm$^{-2}$) is reported, where $V_D$ locates the Dirac point. Resistivities $\rho$ correspond to $\rho$= d•$R_{square}$, where d=3 $10^{-8}$ cm corresponds to a monolayer thickness. Mobilities $\mu$ are determined from the definition $\mu=(n_s e R_{square})^{-1}$ at a charge density of $10^{12}$/cm$^2$ . For 2D and bulk materials bulk values are used (not adjusted for finite size effects).

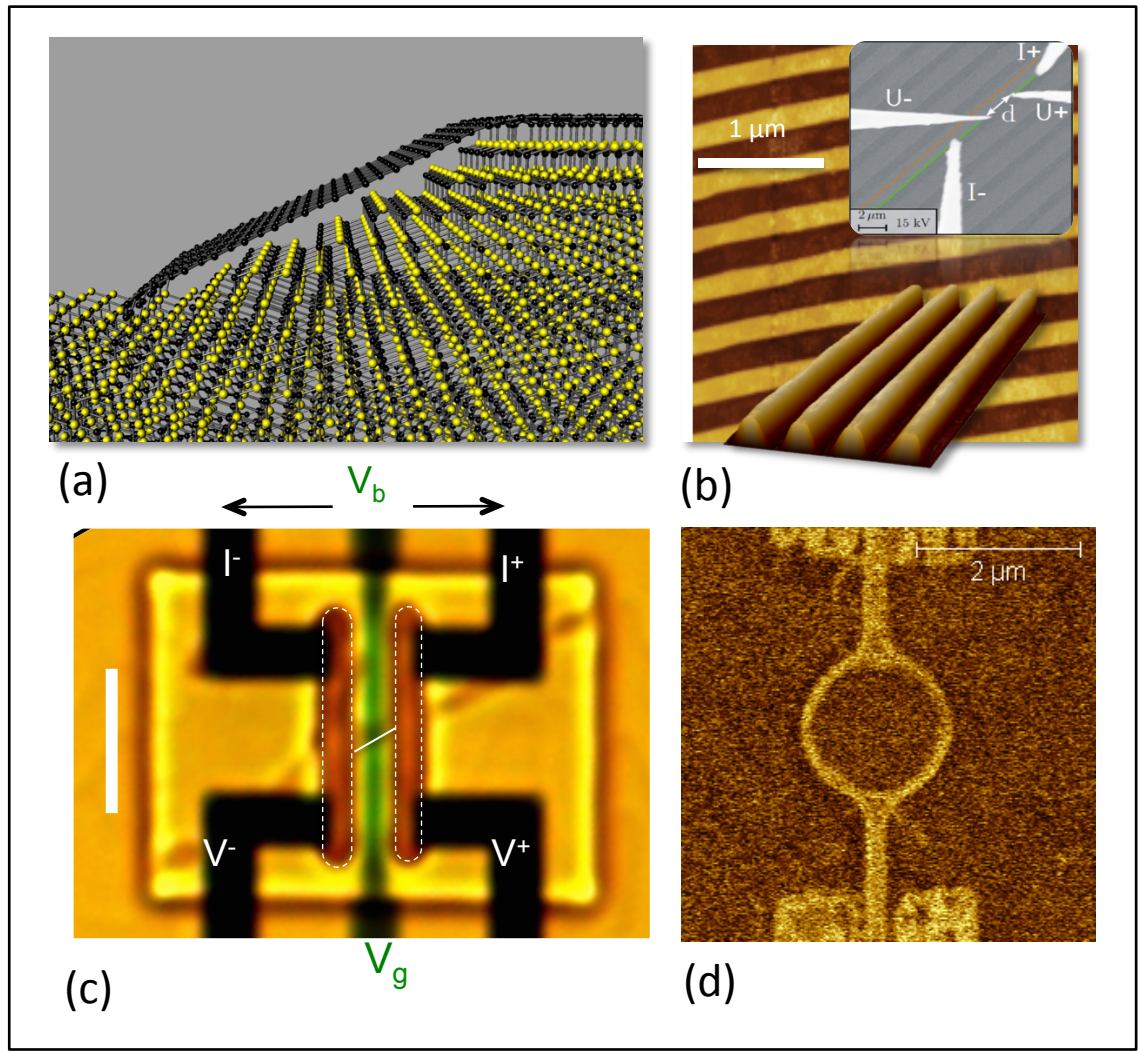

Figure 1

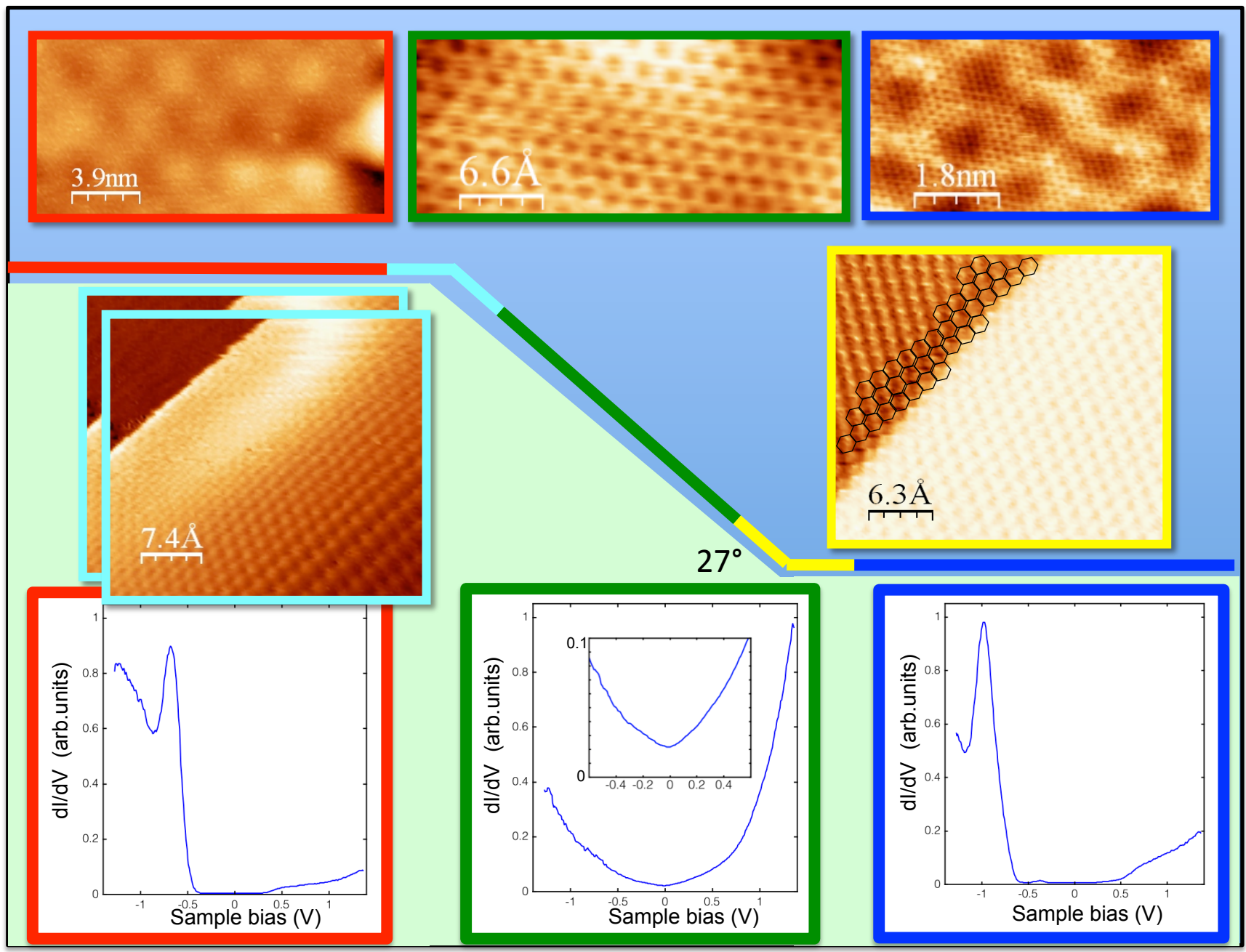

Figure 2

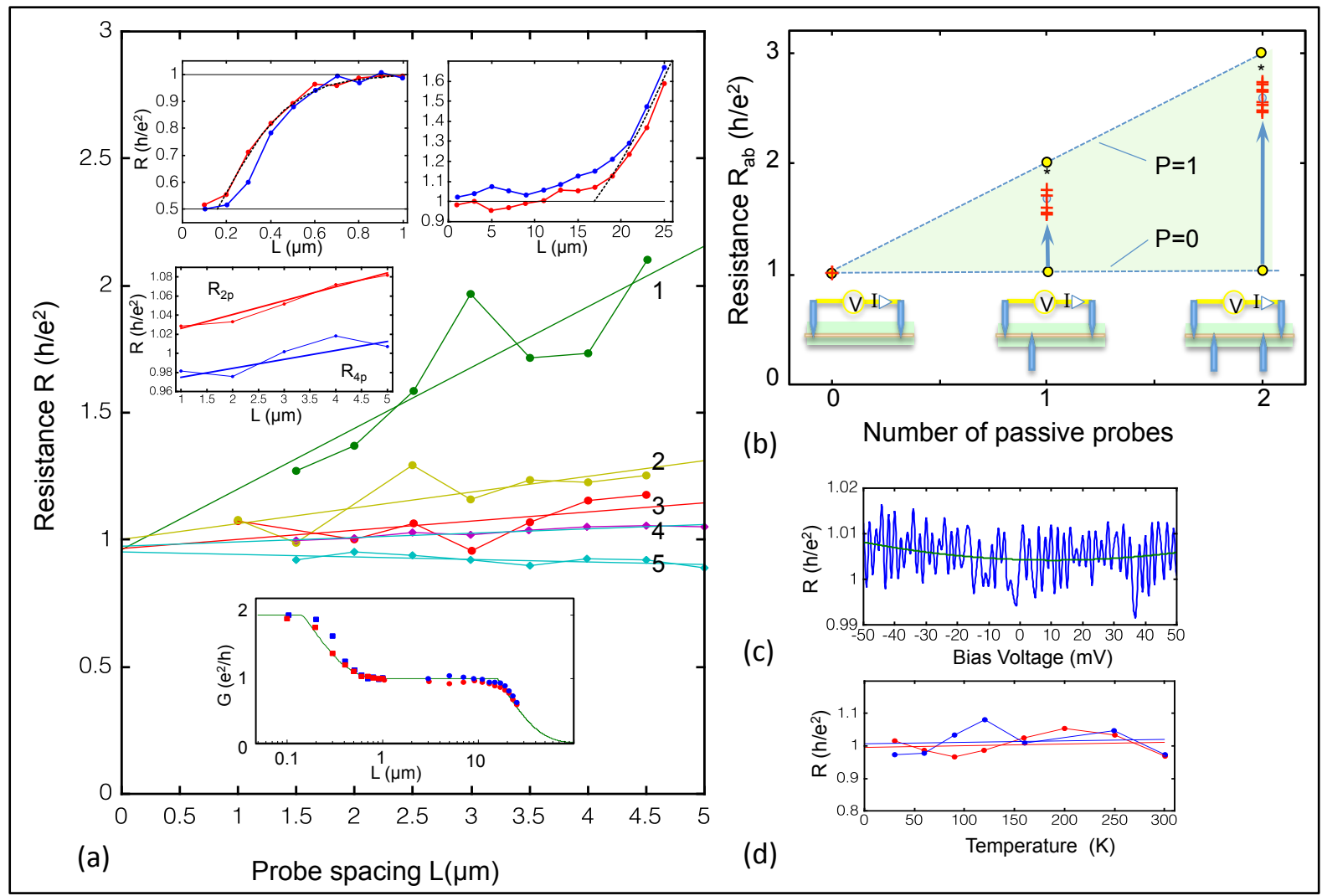

Figure 3

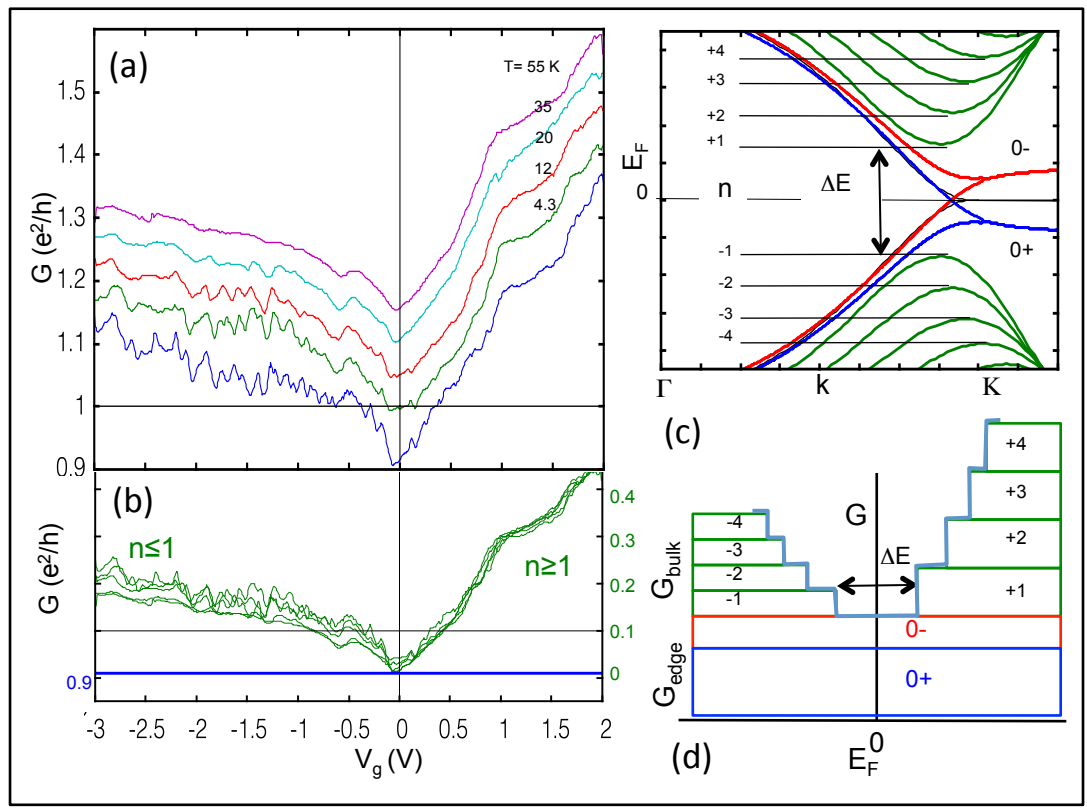

Figure 4

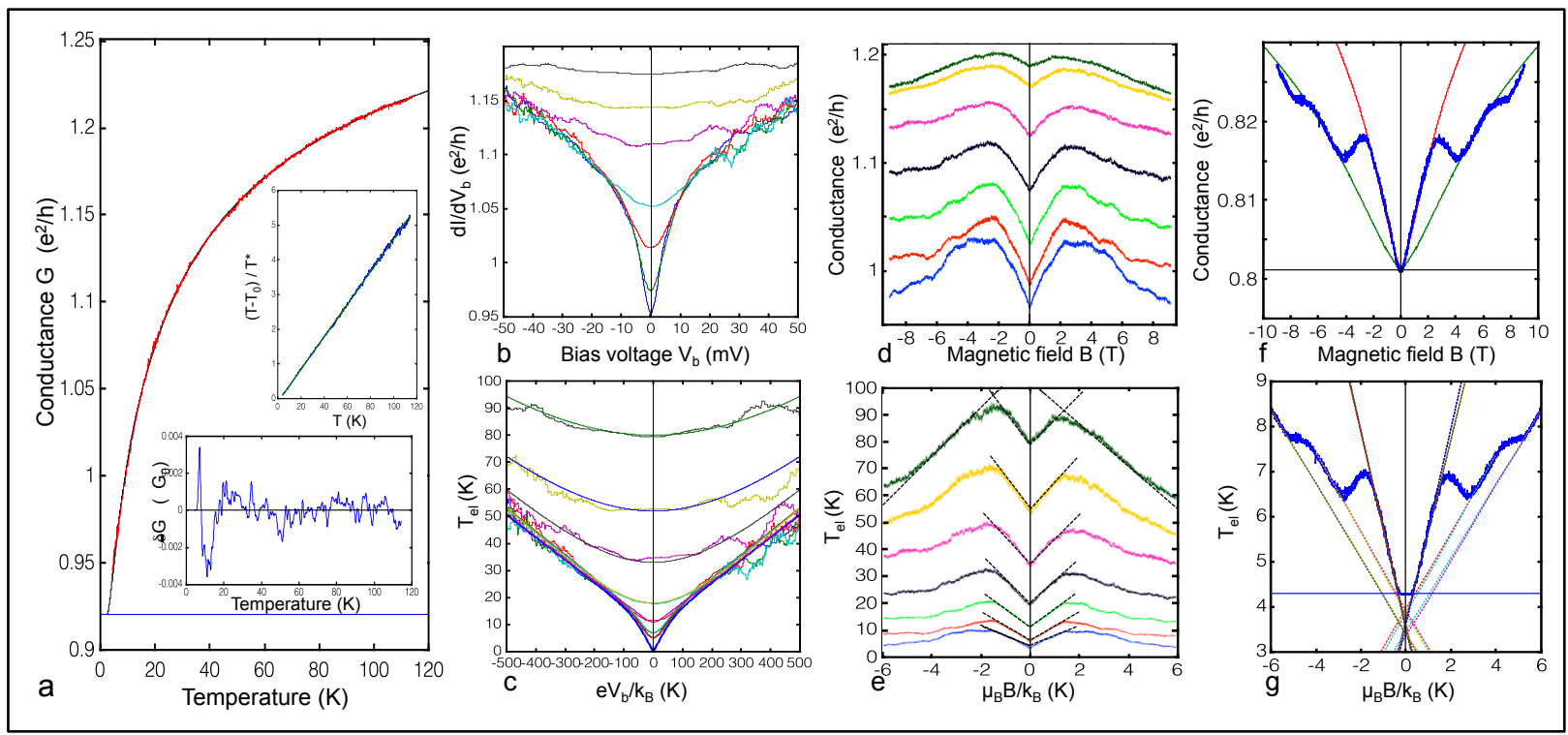

Figure 5

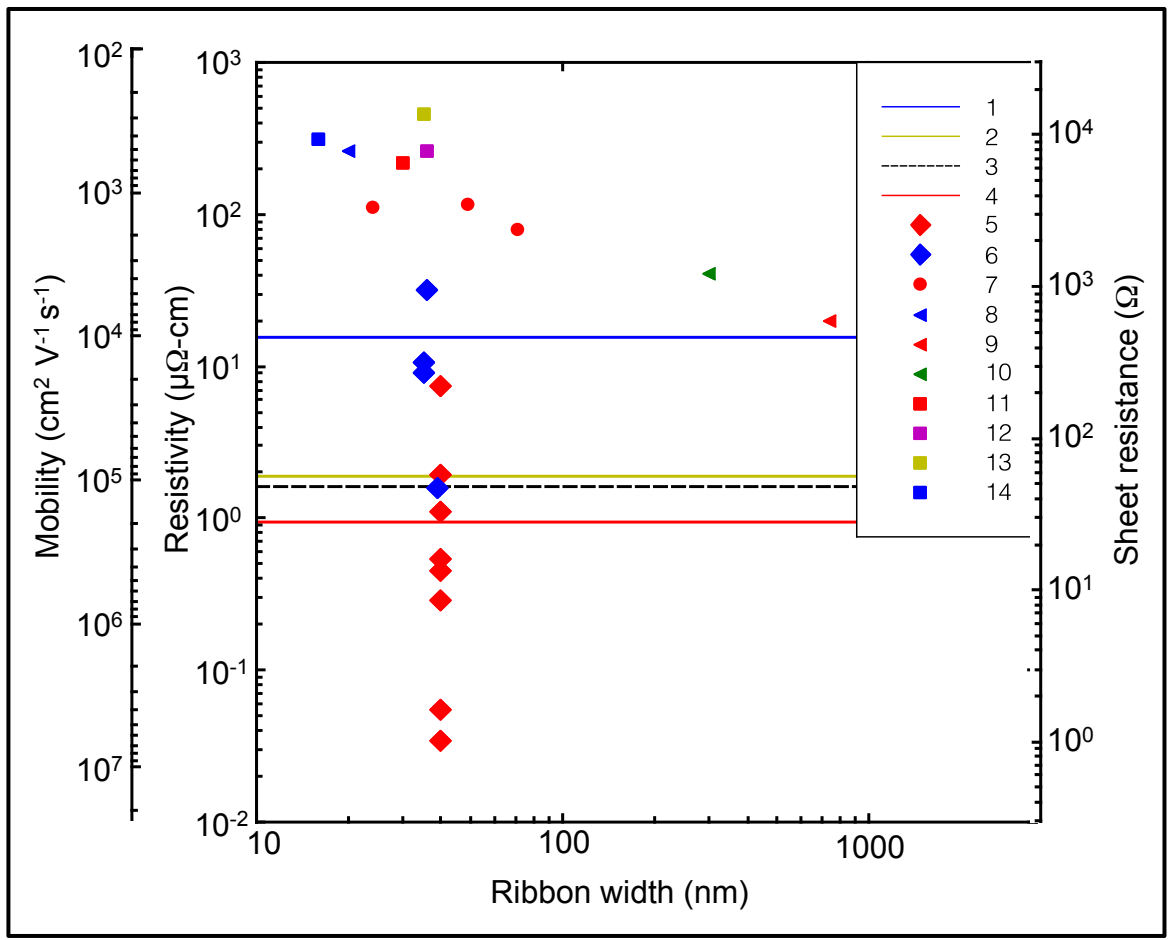

Figure 6

**Supplementary Material**

This section is in three parts. The first part presents Angular Resolved Photoemission Spectroscopy experiments on arrays of epitaxial graphene sidewall ribbons and a comparison of ARPES data with other exfoliated graphene. The second part describes details of the transport analysis presented in the main text, which is brought into context with other graphene ribbon work. We also present details of four samples (A through E) showing that the effects presented in the paper are general. We also provide additional measurements on Sample A, presented in Fig. 4, main text.

The second part describes the preparation and in situ growth and characterization of the graphene nanoribbons in Fig. 3 (main text). It also provides all the data for 50 resistance measurements on various ribbons at 3 different lengths.

    A. Surface characterization: ARPES and STM

    B. Transport analysis

**1. Comparison with exfoliated and other graphene ribbons.**
**2. Graphene Ribbon Sample A**
**3. Graphene ribbon Sample B**
**4. Graphene ribbon Sample C**
**5. Graphene ribbon Sample D**
**6. Length dependence**
**7. The resistance doubling and tripling effect**
**8. Comparison with carbon nanotubes**
**9. FAQs**

    C. In-situ resistance measurements, growth and characterization

**1. Growth of GNRs for in-situ resistance measurements and their -situ characterization**
**2. Characterization by 4-tip STM under SEM**
**3. Additional 4-probe in-situ resistance measurement of GNR**

## A. Surface characterization: ARPES and STM

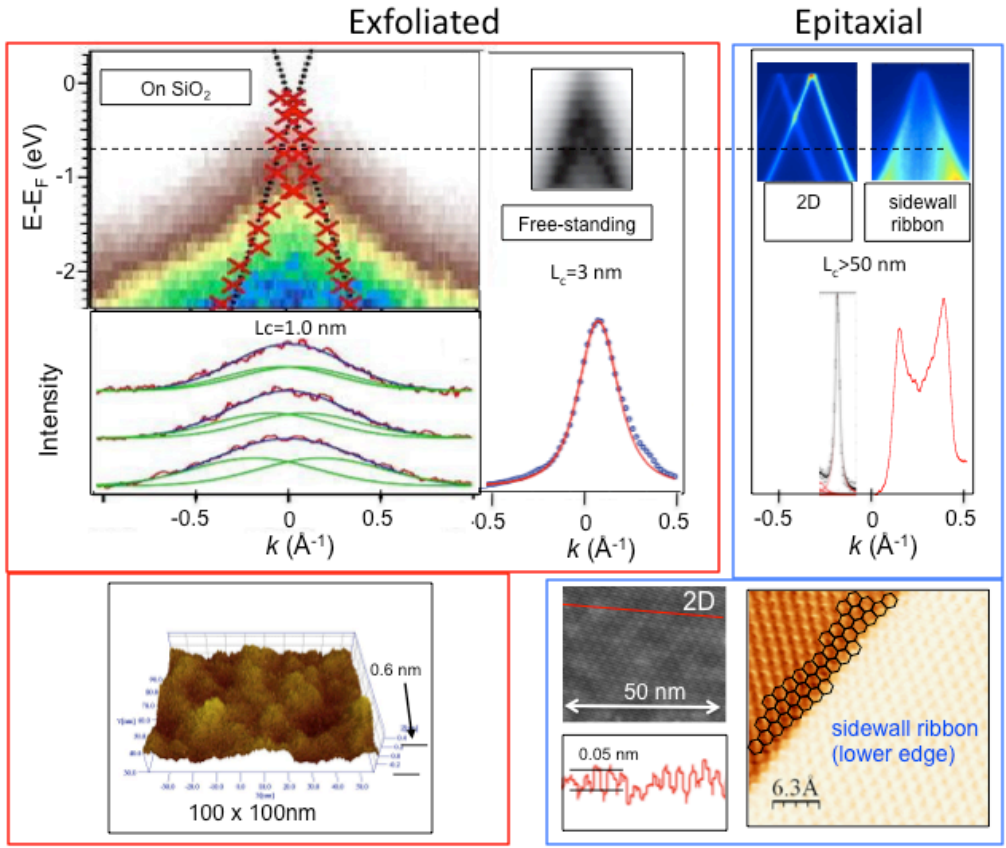

**Figure S1.** Comparison of epitaxial graphene (*blue frame*) and exfoliated graphene (*red frame*) from a surface science perspective. STM topographical images of exfoliated graphene on SiO$_2$ [1] *Bottom left*; and epitaxial graphene[2] (*bottom middle*) and a sidewall ribbon (*bottom right* from Fig. 2). The SiO$_2$ deposited exfoliated material is rough with 0.6 nm corrugations. The epitaxial graphene (both on the 2D, C-face and the sidewall ribbon) clearly shows the honeycomb structure and no substrate induced roughness. 2D epitaxial graphene shows a faint moiré pattern, (not seen in sidewall ribbons.) *Top:* Angle resolved photoelectron spectra (ARPES); Horizontal and vertical scales are the same for all spectra. *Top left*: Exfoliated graphene on SiO$_2$.[3] shows a broad unresolved peak, $\delta k \approx 0.5$ Å$^{-1}$, corresponding to a correlation length $L_c = 2\pi/\delta k = 1$ nm. *Center left*; ARPES of suspended graphene[4], ($\delta k \approx 0.5$ Å$^{-1}$), corresponding to a correlation length of 3 nm (after accounting for broadening due to corrugations). (*Center right*): ARPES of 2D epitaxial graphene on the C-face, the peak width corresponds to the instrument resolution ($\delta k \approx 0.01$ Å$^{-1}$), corresponding to $L_c > 50$ nm. (Left): ARPES of an array of a thousand, 24 nm wide, sidewall ribbons; peak widths correspond to the instrument resolution corresponding to $L_c > 50$ nm parallel to the ribbon axis. Epitaxial graphene exhibits the expected graphene band structure. In comparison, the band structure in all exfoliated graphene samples is severely distorted near the Dirac point. The observed insulating properties of deposited graphene near the Dirac point are a consequence of this disorder. Therefore, exfoliated graphene (free standing or deposited on any substrate) is not reliable as a standard for intrinsic graphene transport properties.

## B. Transport analysis

### 1. Comparison with exfoliated and other graphene ribbons.

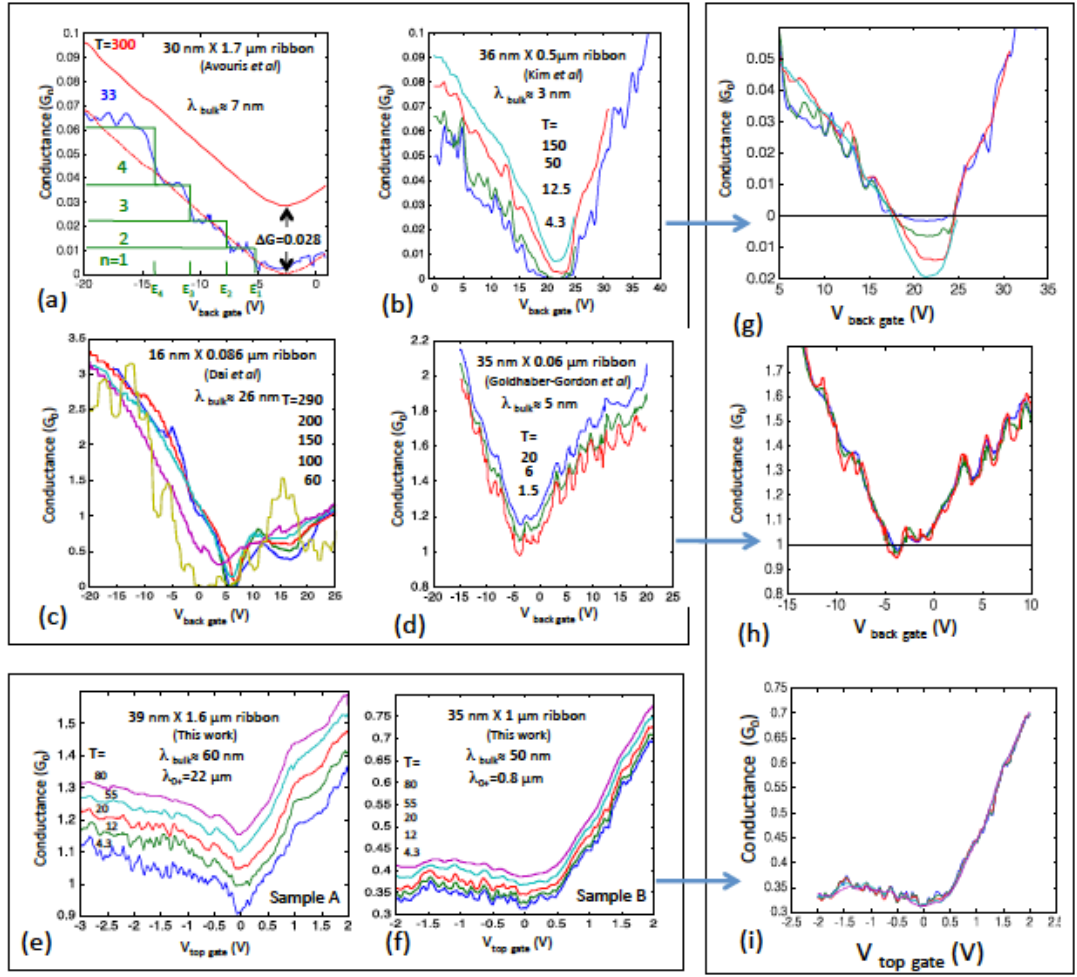

**Figure S2.** Comparison of the conductances of 6 gated ribbon samples measured at several temperatures from the literature, demonstrating temperature independence of n≠0 subbands and temperature dependence of the n=0 subband. (a) 30 nm X 1.7 μm patterned exfoliated graphene ribbon[7]; (b) 36 nm X 0.5 μm patterned exfoliated graphene ribbon [8]; (c) 16 nm X0.086 μm unzipped nanotube [9]; (d) 35 nm X 0.060 μm gate defined exfoliated graphene constriction[10]; (e) 39 nm X 1.6 μm epitaxial sidewall graphene ribbon (Sample A); (f) 35 nm X 1.6 μm epitaxial sidewall graphene ribbon (Sample B); (g) Data from (b) after $G_0(T)$ subtraction; (h) Data from (d) after $G_0(T)$ subtraction; (i) Data from (f) after $G_0(T)$ subtraction. Also note that the mean free paths measured in transport agree well with the correlation lengths that are measured in ARPES (Fig. S1). The poor transport properties of exfoliated graphene ribbons are due to the strong interactions with the disordered substrate and disorder at the interface with the graphene layer. These cause the mobility gap at n=0 in exfoliated graphene ribbons. In contrast, epitaxial graphene sidewall ribbons are ordered (see Fig. S1) and do not have a mobility gap.

Figure S2 shows 6 graphene ribbon samples including 4 samples produced by patterning exfoliated graphene deposited on oxidized silicon wafers (6a-6d). The data were extracted from published work, representing transport properties as a function of temperature and back gate voltages. Samples A and B (6e and 6f) were produced and measured by us. The main point here is the observation that the temperature dependence of the conductivity for $V_g \neq 0$ ($E_F \neq 0$) is represented by a conductivity shift that depends on T only (and not on $V_g$). In the Landauer picture, this shift is consistent with the contribution of the n=0 subbands (i.e. $\mathcal{T}'_0 \to 0$ for T$\to$0) while the transmission coefficients of the n$\neq$0 subbands, i.e. $\mathcal{T}'_{n \neq 0}$ are insensitive to temperature.

- In Fig. S2a a shift of 0.028 $G_0$ causes the T=100 K data to overlap with the T=33 K data.
- In Fig. S2b shifts are applied to produce Fig. S2g.
- Figure S2c is for an opened carbon nanotube ribbon[9] where the n$\neq$0 subbands are reasonably parallel to each other.
- Fig. S2d shows a very short ribbon (a gate-produced constriction[10] ). This constriction is not defined by lithographic patterning of the graphene itself, so that graphene is not damaged and the effective ribbon edges are smooth. Excellent overlap of the data is found by applying uniform G(T) shifts (see Fig. S2h) resulting in a trend that is consistent with Eq. 4, of the main text. The residual conductance is very close to 1$G_0$ and is clearly related to our observations.

An analysis of the scattering lengths shows the following. The back-gate induced charge density is assumed to be equivalent to that of an infinite sheet, yielding an upper limit for the effective scattering lengths along the ribbon. These scattering lengths $\lambda_{bulk}$ are estimated from the measured subband conductivity $\Delta G=G(E_F)-G(E_F=0)$ in units of $G_0$, where, for bulk back gated graphene, $E_F(meV) \approx 31.2\, V_g^{1/2}\, (V)^{1/2}$, so that $\lambda_{bulk} \approx 10.5\, \Delta G\, V_g^{1/2}\, L/W$ where L is the length and W is the width of the ribbon. The scattering lengths determined by this procedure are noted in the figures.

In Fig. S2a, the staircase structure in the 33 K data are due to the opening of successive subbands (as explained by the authors[7]), consistent with the Landauer picture (see main text). The subband indices are shown. Since the step heights $\Delta G_n/G_0$ (=0.011, 0.011, 0.016, 0.022 for n=1-4) correspond to $4\mathcal{T}'_n$, so for small $\lambda_n$, $\lambda_n \approx \mathcal{T}'_n L$=4.7, 4.7, 6.8, 9.3 nm for n=1-4, which agrees with the value determined from the slope $dG/dV_g$. Moreover, the measured energy spacing $E_{n+1}-E_n$=44, 42, 33 meV, agreeing with the energy spacing for a 30 nm zigzag ribbon predicted to be $E_{n+1}-E_n$= 47 meV (for small n).

For comparison, Fig. S2e is a reproduction of Fig. 4a of the main text for Sample A for which $\lambda_{bulk} \approx$ 60 nm and $\lambda_{0+}$ = 22µm Fig. S2f shows a second epitaxial graphene sidewall ribbon, Sample B, (35 nm X 1.06 µm) for which $\lambda_{bulk} \approx$ 50 nm and $\lambda_{0+}$ = 0.8 µm. As shown in Fig. S2i, the $G(V_g)$ all collapse together by applying a $V_g$- independent conductance shift for each temperature. Note that our data shows reproducible fine structure from one temperature to the next, testifying to their high quality.

## 2. Graphene Ribbon Sample A

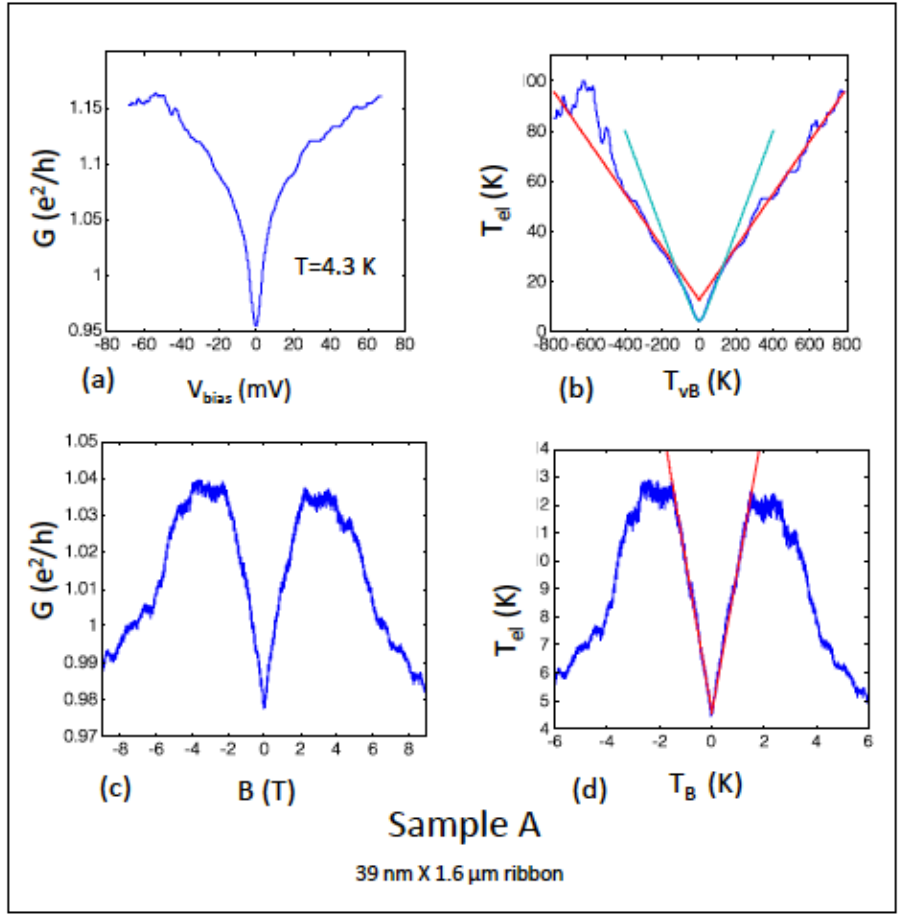

**Figure S3.** Sample A is a 39 nm X 1.6 µm gated sidewall ribbon connected to wide graphene leads. The T=4.3 K conductance versus bias voltage (a); and versus magnetic field (c) are replotted from Fig. 4 main text to emphasize the linearity at low bias voltage (magnetic field, resp); (b) and (d) are the same data as (a) and (c), resp, but plotted in units of electronic temperature $T_{el}$ as a function $T_{vb}= eV_b/k_B(K)$ and $T_B= \mu_B B/k_B$., resp.,(see Fig.4)

An extension to the detailed analysis already presented in the main text concerning sample A, highlighting the T=4.3 K data is presented here. The conductance G of this sample is given by $G(T)=\alpha(1+1/2 \exp-\theta^{1/2})$, where $\alpha=0.922$; $\theta =T^*/(T-T_0)$ with $T^*=21.13$ K, (compared with $T^*= 19.5$ K from $T^*=4\hbar c^*/Lk_B$ (Eq. 4b, L=1.6 µm) and $T_0=2.2$ K, This equation is algebraically inverted to determine the electronic temperature $T_{el}$ from the conductance: $T_{el}(G)=T^*(\log(2G/\alpha -2)^{-2}+T_0$. Likewise, the bias voltage is converted to $T_{vb}=eV_b/k_B$ and the magnetic field is converted to $T_B=\mu_B B/k_B$. In this way Figures S3b and S3d are generated (from S3a and S3c). The slopes in S2d correspond to the inverse electronic heat capacity $\nu$ in units of $k_B$ (as defined in the main text) slopes in S3c correspond to magnetic moments (in units of $\mu_B$), Note the sharp V shaped dip in the magnetic data (with slopes corresponding to 5.6 $\mu_B$), which is typical for all low temperature, low field graphene ribbons. The lines corresponds to Eq. 4d (main text) with $T_{el}=T_1+\mu|B|$ with $\mu$=5.6 $\mu_B$ and $T_1$=4.3K, corresponding to the sample temperature. The downturn at $T_B=\pm 2$K, is seen in most graphene ribbon, 2D graphene and nanotube data. The hyperbolic fit to Fig. S3b, corresponds to $T_{el}=\sqrt{[(T_{vb}/\nu)^2 + T_1^2]}$ with $\nu$=5 and $T_1$= 4.3 K (i.e. the sample temperature) as in Eq. 4c (main text).

### 3. Graphene ribbon Sample B.

Sample B is a graphene ribbon (1.06 μm X 35 nm) sample similar to sample A in design. The main features observed in sample A are also observed here. The temperature dependent conductivity corresponds to $\alpha$= 0.311 and T*=29 K. Note that the predicted T* according to Eq. 4b is 29K, in excellent agreement with the measured value (despite the significantly reduced value of $\alpha$).

The G versus $V_g$ for several temperatures is similar to Sample A, Fig. 4a (main text)**.** Note that all of the data collapse onto a common curve after subtraction of a gate voltage independent conductance (temperature dependent) as was seen in Sample A as well (Fig. 4a). The implications of this are discussed in the main text (see also discussion of Fig. S2).

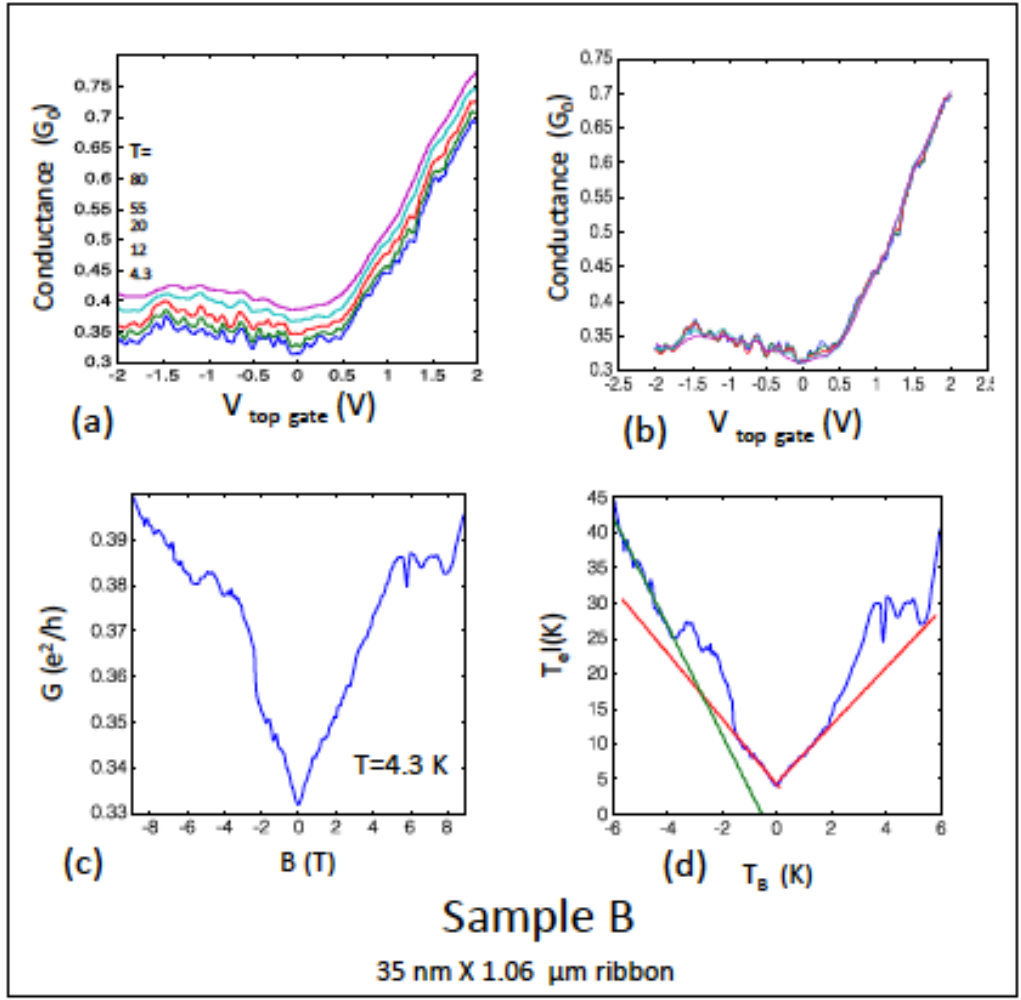

**Figure S4** Properties of Sample B (35 nm X 1.06 μm sidewall ribbon supplied with a top gate). (a) G versus $V_g$ for various temperatures. (b) Collapse of the data in (a) by subtraction of a $V_{bias}$ -constant G(T) for each temperatures T. (c) Conductance as a function of magnetic field for T=4.3 K. (c) $T_{el}$ versus $T_B$ (see discussion of Fig S3 or main text). Slopes correspond to 4.3 $\mu_B$ (red) and 7.7 $\mu_B$ (green)

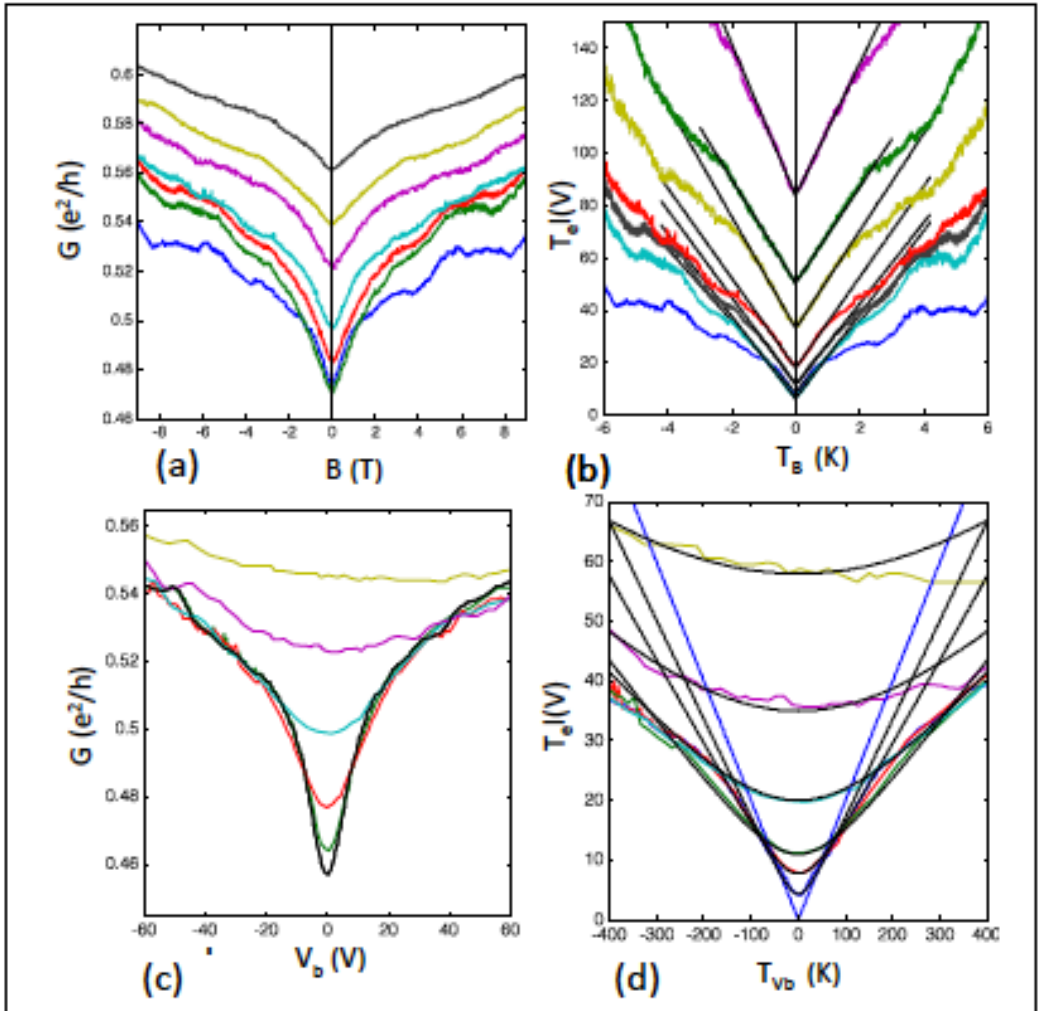

**Figure S5.** Graphene ribbon Sample C, L= 1 μm, W=39 nm ; T*=49 K, $\alpha$= 0.454 (a) Conductance versus magnetic field, measured at T=4K; 7K; 12K; 20K; 35K; 55K; 120K; and 180 K showing characteristic dip at B=0. (b) Data from (a) after conversion to $T_{el}$ and $T_B$, showing linear response agreeing with Eq. 4d. Magnetic moments correspond to 14 $\mu_B$; 16 $\mu_B$; 16 $\mu_B$; 17 $\mu_B$; 20 $\mu_B$; 20 $\mu_B$; and 27 $\mu_B$. (c) Bias voltage response for T= 4.3 K; 7 K; 12 K; 20 K; 35 K; 55 K and 120 K. (d) Data from (c) after conversion to $T_{el}$ and $T_{vb}$ . Hyperbolic response agrees with Eq. 4c with ν= 5; 6; 7; 9.5; 9.5; 11; 12; 12.

## 4. Graphene ribbon Sample C

The magnetic field and bias voltage dependence of graphene ribbon Sample C (measured from 4.3 K to 180 K) shows the characteristic features of graphene ribbons (Fig. S5). After converting G(B) to $T_{eff}$ and B to $T_B$ (like in the previous examples), the highly curved magnetic field response shows the typical behavior of Eq. 4d. However the magnetic moments obtained from the slopes range from 14 $\mu_B$ to 27 $\mu_B$; that are a factor of 2 greater than typically observed. The bias voltage dependence shows the hyperbolic response (Fig. S5 d) as predicted in Eq. 4c, with $\nu$= 5 at 4.3 K, typical for graphene ribbons. The (approximately factor of 2) slope change at $T_{el} \approx$ 15K is also typically seen, as reported in the main text.

**5. Graphene Ribbon Sample D**
Sample D is a sidewall ribbon 36 nm X 0.37 µm. From G(T), $\alpha$= 0.628 and T* =87 K The value of T* determined from Eq. 4b is 85 K that agrees very well with the measured value. Conductance measurements were made on this graphene ribbon, as shown in Fig. S6a, using a conducting AFM tip at room temperature in ambient conditions. The left wide graphene pad was connected to ground. The contact resistance, measured by placing the tip on the left graphene pad was subtracted. The resulting conductance versus tip position is plotted in Fig. S6b, showing that the conductance decreases from about 2 $G_0$ to about 0.9 $G_0$ with increasing tip to pad distance L. The decrease is similar to that observed in Fig. 3a for Sample A, (main text) as indicated by the theoretical curve (green line). The curve is identical to that in Fig. 3a where the exponential decrease is given by: G=$G_0$exp(1-L/L*) for L>L*, where L*=hc*/$k_B$T = 160 nm. While the correspondence is not nearly as good as in Fig. 3a, it is nevertheless consistent with the measurement.

The magnetic field and bias voltage dependence of this sample are typical for graphene ribbons and agree with Eq. 4c,d (simulated curves are from Eq. 4). Measurements were made for temperature ranging from 4.3K to 180 K. The curved response of the raw data (Fig. S6c) converts to the typical V shape with a magnetic moment 5.4 µB at T=4.3 K as typically seen in graphene ribbons. The bias voltage response is also typical. The bias voltage dip converts to hyperbola given by Eq. 4c; with $\nu$= 7 for T=4.3 K (see the figure caption of Fig. S6 for the other values).

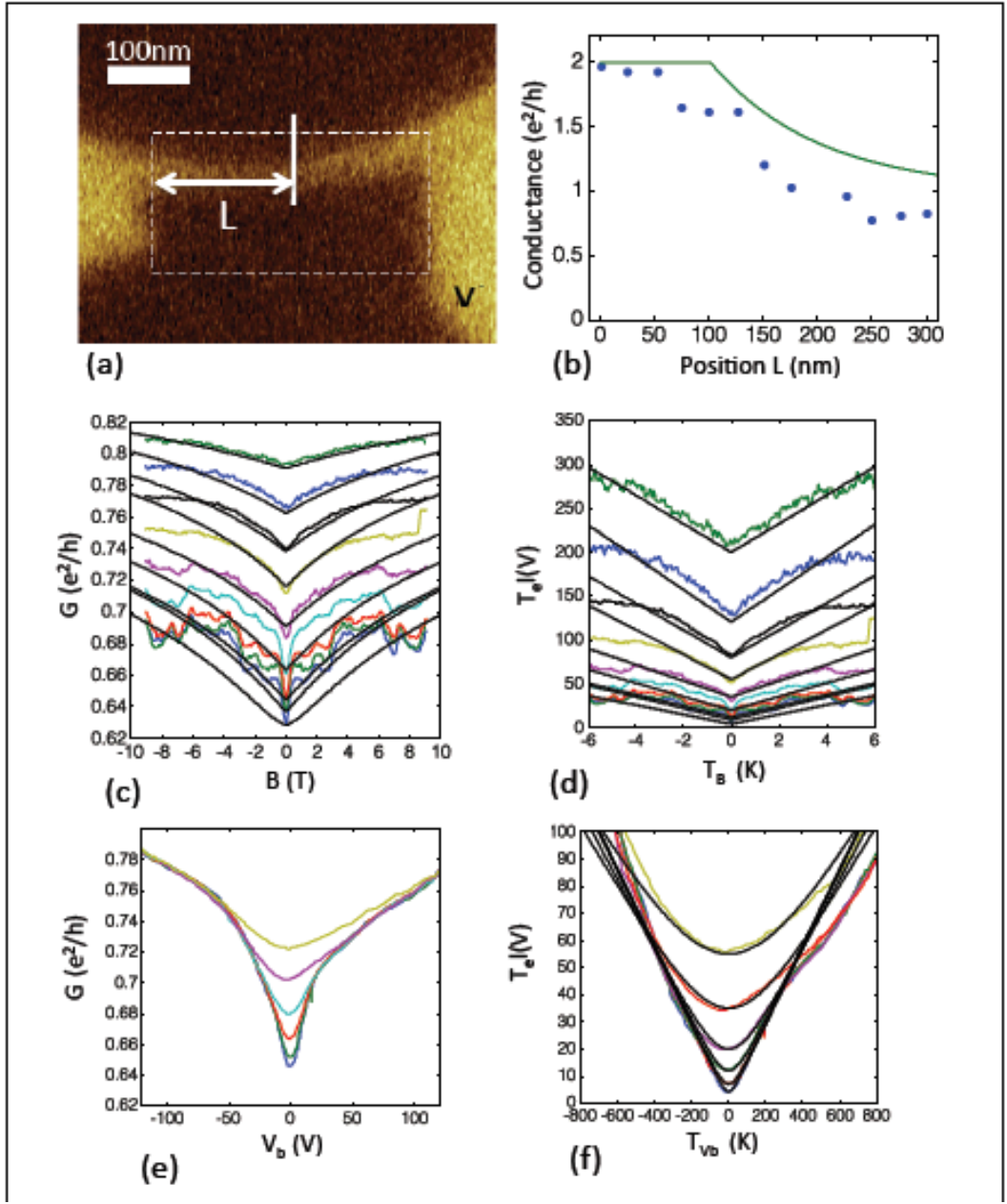

**Figure S6.** Transport properties of Sample D, a 36 nm X 0.37 μm sidewall ribbon. (a) Electrostatic force microscopy image. (b) Scanning probe conductance measurements using conducting AFM tip showing uniform conductance decrease from G=2 G$_0$ to G=0.8 G$_0$. Solid line corresponds to Eq, 3a, (main text). (c) Magnetic field dependence for T= 4.3, 7, 12, 20, 35, 55, 80, 120, and 180 K. Fits correspond to Eq. 4a with α=0.628 and measured T*= 87K (which agrees very will with the predicted T*=85 K from Eq. 4b at the corresponding temperatures. The magnetic moments correspond to μ/μ$_B$=5.4, 6.5, 6.4, 7.7, 9.2 ,14, 15, 18, and 16 as typically found in graphene ribbons. (d) Same data as in (c) plotted in terms of T$_{el}$ and T$_B$, explicitly showing the predicted behavior (Eq. 4c). (e) Bias voltage dependence for T= 4.3K, 7K, 12K, 20K and 55K. (f) Data in (e) plotted in terms of $T_{el}$ and $T_{Vb}$ using Eq. 4c, for ν=7.0, 7.0, 7.0, 7.5, 8.3, and 8.1; values that are typical for graphene ribbons.

## 6. Length dependence

The length dependence of the 0- subband for T=300 K (see Fig. 3a, main text) is plotted on a logarithmic scale (Fig. S7), which brings out its exponential behavior more clearly for L>L*= $hc^*/k_B T$=160 nm,. For both ribbons the conductance is approximately $G=G_0 \exp(1-L/L^*)$ for L>L* and $G=G_0$ for L<L* as explained in the main text.

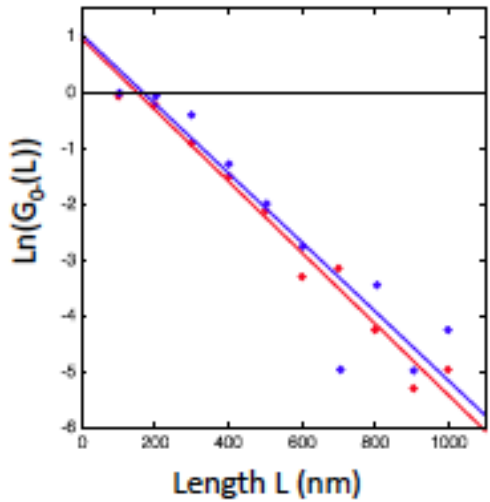

**Figure S7.** Log $G_{0-}$ versus length for the two short ribbon segments in Fig. 3a (Main text) measured at room temperature.

## 7. Comparison with carbon nanotubes

As shown below the response of carbon nanotubes is remarkably similar to that of graphene ribbons. However the overall conductances are increased by a factor of 2. As was shown 15 years ago (see Ref. 1, main text), carbon nanotubes are ballistic conductors at room temperature and the conductance is (nominally) $2e^2/h$, in contrast to the theoretical prediction of 4 $e^2/h$. In that respect, similar to the factor of 2 discrepancy observed in graphene ribbons. From thorough transport measurements on (multiwall) carbon nanotubes, Schönenberger et al. (Ref. 34 main text) concluded that (1) Carbon nanotubes are quasi-ballistic with mean free paths on the order of 20 nm. (2) Their magnetic response shows a conductance dip that can be modeled with a multiparameter fit to standard weak localization theory. (3) Their bias voltage response demonstrates Luttinger liquid behavior. (4) The temperature response is complex and shows evidence of localization and no evidence for room temperature micron scale ballistic conductance.

In light of our measurements on graphene ribbons we reexamined the data from which these conclusions were drawn and found carbon nanotubes behaved essentially identically to graphene ribbons as shown in Fig. S8 in all details. Specifically, the temperature dependence is consistent with Eq. 4a with $T^*=49.9K$ (corresponding to L=620 nm, from Eq. 4b, compared with the measured L=350 nm.) $T_0$= 2.0 K. However, the factor of 1/2 in Eq. 4a is replaced by a factor of 3. The conductance G(T) saturates at G=2 $\alpha(2e^2/h)$ at T≈50K (Fig. S8b), a factor 2 higher that the short ribbons in Fig. 3a.

Like for graphene ribbons, the bias voltage dependence is found to follow Eq. 4c, with ν=5 (see Fig. S8c) for small bias voltages and about twice that for large bias voltages. The magnetic field dependence shows the sharp characteristic dip at B=0 that corresponds to Eq. 4d with µ= 5 $\mu_B$. The V shape is interrupted for |B|≥2T showing complex behavior at higher field, as seen in all graphene ribbon samples. Consequently, the transport properties carbon nanotubes, as well as underlying mechanisms are certainly similar to graphene ribbons. Nanotubes are certainly two component ballistic conductors as well.

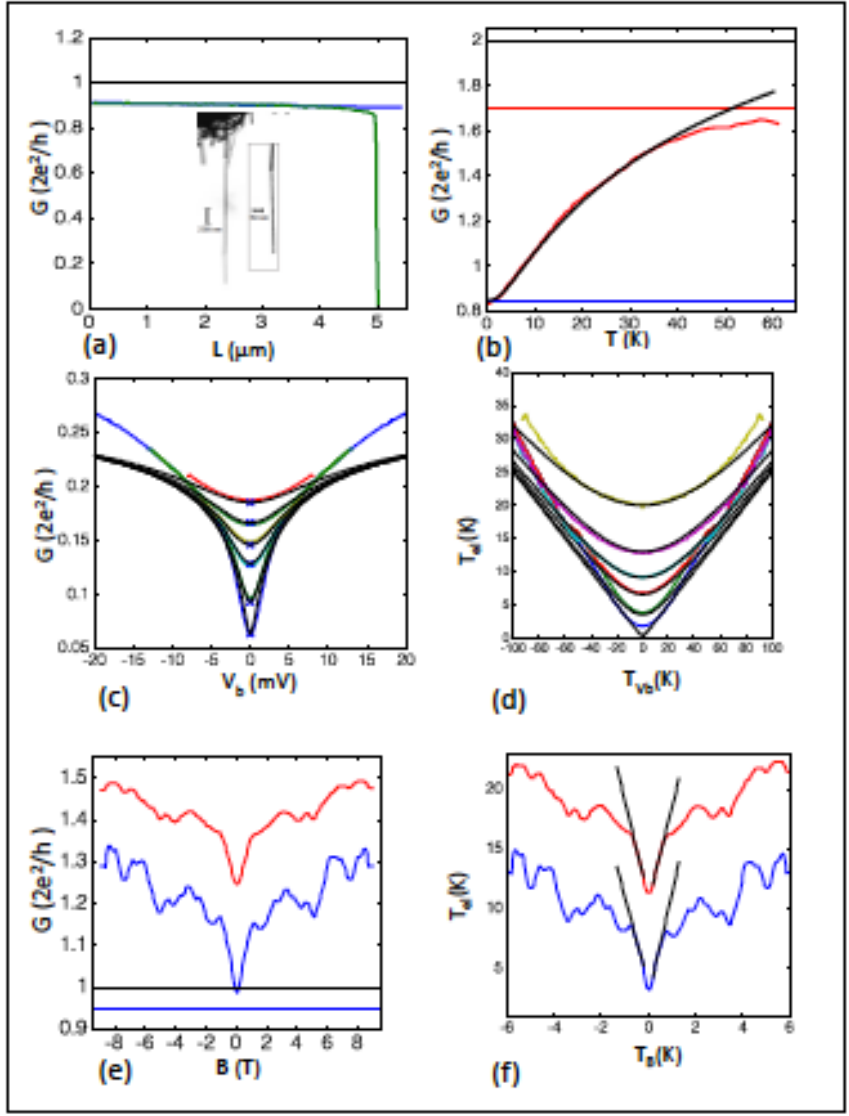

**Figure S8** Reexamination of transport properties of multiwall carbon nanotubes (a) Room temperature transport as measured by Frank et al (Ref. 1) showing $2e^2/h$ ballistic transport indicating two channels. (b-f) Measurements by Schönenberger et al[11]. (b) Measured temperature dependence of the conductance, corresponding to $\alpha= 0.90$, $T^*=49.9$. Note the saturation at $2\alpha$. (c) Measured bias voltage versus temperature with superimposed calculations following Eq. 4. The fits correspond to $\nu=4$, very close to $\nu=5$ observed in graphene ribbons. (d) Data of (c) converted to $T_{el}$ and $T_{vb}$ showing hyperbolic behavior consistent with Eq. 4c. (e). Magnetic field dependence for T=2.5 K and T=12 K, showing typical dip at B=0; (f) Same data as in (e) plotted as a function of $T_{el}$ and $T_B$, with superimposed simulation according to Eq. 4c,d, with $\mu= 5\mu_B$, as for graphene ribbons. As for ribbons, the behavior becomes complex for |B|>2T.

# 8. Explanation of the resistance doubling and tripling effect

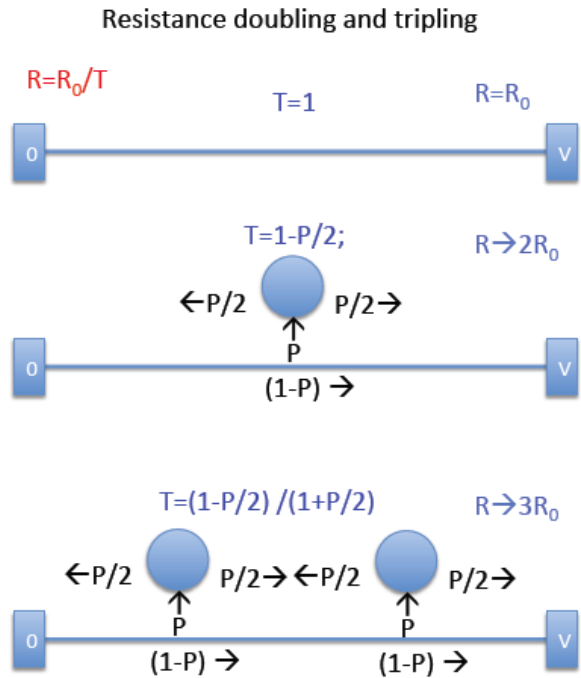

**Figure S9.** Ballistic wire connected to contacts (left and right) as in Fig. S11. (Top) Undisturbed wire. (Middle) wire with one thermalizing probe. (Right) wire with two thermalizing probe.

Fig. S9 shows how invasive probes affect a current flow. For a ballistic one-dimensional non-degenerate wire, the conductance $G=G_0T$. If the flow is undisturbed then $T=1$ and $G=G_0=e^2/h$ (Fig. S11, top). For a more complete discussion, see[12,13].

However an invasive probe will alter the flow. P is the probability that a charge carrier moving past the probe will enter it. After thermalization in the probe, it exits it with equal probability going either to the right or to the left. As indicated in Fig. S9 the total forward probability is (1-P)+P/2 (the reflected probability is P/2). Hence, $T=1-P/2$ and $G=(1-P/2)G_0$. If P=1 (every electron enters and exits the probe) then $G=G_0/2$, (the resistance is doubled: $R=2R_0$). In fact this is what should be expected, since a perfectly invasive probe simply divides the ballistic wire into two ballistic wires.

If two invasive probes are inserted (Fig. S9 bottom) then the situation is slightly more complicated, since the backward scattered charges from probe 2 (for example) will scatter from probe 1 etc. This results in a geometric series that is easily summed to give $T=(1-P/2)/(1+P/2)$; Hence, for P=1, T=1/3. R→3$R_0$

## 9. FAQs : Alternative explanations suggested by referees and others.

*The entire surface is graphitized with a diffusive conducting layer giving rise to high conductivities; the observed conductance quantization is simply fortuitous.*

This morphology is contradicted by numerous experiments, see for example Fig. 2, and extensive Raman and EFM characterization (see for instance 1).
Resistances between adjacent ribbons are at least 300 kΩ (and more than 30MΩ for the low-temperature measurements), at least an order of magnitude greater than along a ribbon.
This interpretation cannot explain the dependence on the probe separation.
It cannot explain the resistance doubling and tripling effect.
The probability P that measured values to fall within 20% of each other in more than 50 samples (assuming a random distribution with a dispersion of a factor of 2, which is very conservative) is $P=10^{-50}$. Hence it is impossible.

*The graphene is multilayered making the conductivities large.*

ARPES measurements show a monolayer. Moreover, it would require at least 100 highly doped layers to attain the observed resistivities (dR/dL), which is physically impossible. Also, this cannot explain the consistent 1 ($h/e^2$) contact resistance (nor any of the other observed effects).

*The ribbons are diffusive and the resistance doubling and tripling effect is caused by multiple side-by-side ribbons that are broken when they are touched by the probe and reunite when the probe is removed.*

It is inconceivable that ribbons break and reform when touched by a probe. In order for several ribbons to consist of three ribbons in parallel to produce this effect would require an impossible combination of parallel diffusive ribbons.

*The magnetic field dependence and temperature dependence is due to weak localization.*

The fit to Eq. 4 is exceptionally good and requires a minimal number of parameters, each of which are well defined and consistent from one ribbon to the next. Neither the temperature dependence nor the magnetic field dependence nor the length dependence can be reproduced over the same range with using standard weak localization theory with accepted expressions for $L_\phi$ and $L_m$ (see Ref. Schönenberger (ref 38 in text) or Beenakker (ref 22)). For example, for carbon nanotubes, two different power laws are need for $L_\phi$ for 2 K ≤T≤10 K and 10 K<T< 50 K an(see Ref. Schonenberger).

*The bias voltage dependence is due to the well-known zero bias anomaly.*

Attempts to fit the bias voltage dependence with standard (Luttinger liquid power law) approaches failed to reproduce the observed bias voltage dependence over any reasonable range.

*Properties measured on epitaxial graphene are not relevant because substrate interactions are large; it is not "real graphene" but rather a "graphene-like material".*

Properties within 1 meV from charge neutrality are reliably attributed to graphene as experimentally demonstrated (see main text). Substrate interactions and strain-induced effects measured in ARPES, STM and in transport are very small and much smaller than in deposited and suspended graphene as clear from the (scientific) literature. References are given in the text. See also Figs. S1 and S2. The pervasive misperception of the role of substrate interactions in epitaxial graphene (experimentally shown to be small in the 1990's) and in exfoliated graphene (explicitly shown to be large in 2007) is inexplicable*. Especially since (in contrast to epitaxial graphene) the ambient conditions under which exfoliated graphene is produced, makes it prone to extreme contamination, charge disorder, random stress, and structural disorder, all which are observed and all of which are known affect transport.

*For example, in the abstract in Phys. Rev. B 58 16396B, 1998 concerning epitaxial graphene on SiC (cited 350 times). Forbeaux et al write:
 *"The observation of unshifted π* states, which reveals a very weak interaction with the substrate, is consistent with the growth of a van der Waals heteroepitaxial graphite lattice on top of silicon carbide."*
This observation has been confirmed in many observations since then. Nevertheless, from the outset, leaders of the exfoliated graphene community have insisted that substrate interactions in epitaxial graphene are large compared with those in exfoliated graphene (without providing scientific evidence), to support claims that epitaxial graphene is not "real graphene" but a graphene-like material. This misperception (and others like it) continues to be parroted in prominent reviews.

## C. In-situ resistance measurements, growth and characterization

### 1. Growth of GNRs for the in-situ resistance measurements and their ex-situ characterization

The GNR structures for the in-situ resistance measurements were grown in Hannover selectively by sublimation epitaxy on MESA-structured 6H-SiC(0001) surfaces [14]. Before, 1 μm wide line structures were generated by optical lithography (UV-light, 286nm) and reactive ion etching (RIE, $SF_6$ and $O_2$ ratio 20:7) onto the 6H-SiC(0001) surface (nitrogen doped, $10^{18}$ cm$^{-3}$). The optical mask was aligned such that the trench structures run along the [1-100] direction, i.e. the zig-zag direction for graphene grown epitaxially on Si-terminated SiC(0001). The anisotropic etching and suitable etching rates of around 0.3 nm/sec allows us to fabricate defined terraces and trench MESA structures. By thermal annealing (DC-heating in an Ar atmosphere of $4\times10^{-5}$ mbar, sample clamped by graphite contatcs) of this structure clean and well-ordered crystal facets are forming around 1420 K [14,15]. Further annealing to 1570 K results in growth of extended graphene nanoribbons on these facets [14,16] as sketched in Fig. S10. The formation of well orientated graphene nanoribbon structures has been proven recently by LEED and ARPES measurements [6].

In this study ribbons down to 40 nm in width were obtained by using MESA trench structures of 20 nm in depth. Electrical measurements on these nanostructures were performed with a 4-tip STM/SEM system (Fa. Omicron nanoprobe). Details are reported below. Before transfer of these structures into the 4-tip STM SEM system for further processing and electrical characterization the overall quality of the ribbons has been checked by AFM and EFM (see Fig. S10d,e). The line scan of the AFM demonstrates nicely the accuracy of the etching process. The local change of the work function upon formation of graphene at the sidewalls has been monitored using electrostatic force microscopy (EFM) as shown in Fig. S10e. After deconvolution of the AFM tip radius the full widths of the EFM-peaks located at the facet sites represent almost the width of the GNRs.

### 2. Characterization by 4-tip STM under SEM

The GNRs have been characterized in-situ by means of electrical transport measurements using a 4-tip STM/SEM system (Nanoprobe system, Fa. Omicron). The system operates at a base pressure of $10^{-8}$ Pa and by cooling with Liq-He, temperatures down to 30 K can be obtained. By means of the in-situ high-resolution SEM (<4 nm) the tungsten-tips can be navigated to desired positions above the nanostructures and approached individually to the surface via feedback control approach mechanisms. The transport measurements in this study were performed usually in the following way:

Prior to measurements on the GNRs, the W-tips (NaOH-etched) have been "calibrated" after installation. By means of sheet conductance in 2d-graphene on SiC we have ensured that the tips are mechanically stable with a geometrically small (20-40nm radius) and metallic apex structures. As mentioned above with the help of SEM the tips have been navigated to individual ribbons and placed *above* the ribbon in a collinear arrangement with well-defined equal inter-probe spacings d. Each tip has been approached via a feedback controlled loop into a tunneling contact at its desired position (set point +2V, 1nA). At first hereafter, the feedback was switched off and the tips approached via calibrated piezo-elements pressing *on top* of the ribbons for the final transport measurements. After each measurement the ribbons have been carefully checked by SEM in order to exclude tip-induced changes to GNRs (and to tips).

The ex-situ processed GNR-samples have been annealed (600°C) in-situ in order to remove organic contaminations adsorbed during transfer. Furthermore, high temperature annealing (>1300°C) is possible in this system as well and has been used occasionally to improve further the quality, i.e. the mean free paths (the $\lambda_{0+}$ - see main text – Fig.3), of the ribbons.

The selective growth of graphene nanoribbons (GNR) is demonstrated here by lateral four-probe measurements and local tunneling spectroscopy (STS). A typical tip assembly is shown in Fig. S11a). The resistances were calculated from I(V) curves in a current range of +/- 1µA (cf. Fig. S11b). Most noticeably, the resistance measured on the sidewall is around 26 kΩ and almost by a factor of 20 lower compared to collinear transport measurements on the terraces (dashed line in Fig. S10a) or valleys of the MESAs. The resistances on these areas are finite, possibly due to the SiC doping, but can be well discriminated from the resistances measured at the side walls.

Local spectroscopy (STS), performed with the tip moved by the high resolution scanner in the system, has been used in addition to determine the chemical potential of the GNRs. As seen in Fig. S11c the GNR is slightly p-doped ($E_F$ = 150 meV below $E_D$) in agreement with ARPES measurements on GNR array structures processed in a similar manner[6]. The fact that the chemical potential coincides de-facto with the Dirac point ($E_D$) ensures, that only low lying subbands are occupied with electrons (see discussion below).

The STS spectrum taken on the terrace structure shows in contrast a gap of more than 1eV at $E_F$. This supports our conclusions that spatially extended graphene nanostructures are formed exclusively at the step edges of the MESA structures. The onset of the current seen in the STS spectrum in the negative bias regime, which probes the occupied surface states, correlates nicely with ARPES data taken solely on buffer layer structures[17]. Please note, the STS spectra were taken by positing the tips with radii of 20-40 nm roughly above center of the GNR, thus the spectrum represents basically an average of the electronic states across the ribbon structure.

**3. Additional 4-probe in-situ resistance measurement of GNR**

The transport properties for various GNR structures have been systematically investigated in-situ with respect to the number of contact probes, contact separation and sample temperature. As outlined in the main text, the $0_-$ channel start to localize for distances ≥ 150nm. Consequently, to probe both channels requires extremely small contact spacing (<100nm), which is experimentally very demanding. In contrast the $0_+$ channel that shows a $e^2/h$ behavior over long distance is easily measured using larger spacing. .

The robustness of he ballistic behavior over long distance of the $0_+$ channel has been verified for many different ribbons. In total 50 different GNRs have been probed for three contact spacing in the intermediate length regime (L=500nm, 1.5µm, and 5µm) and various temperatures (32K, 78K, 120K, 298K). All ribbons were located within an area of 100 x 100 µm². Their values (absolute and relative to $G_0=e^2/h$ ) are listed in Table 1 below and visualized by the histogram in Fig. S12. Most ribbons show a $e^2/h$ conductance and the variance can be correlated with the probe spacing: Higher (lower) conductance values correspond to shorter (larger) probe spacing due to contributions of the 0. The variation within each spacing regime is attributed to slightly different mean free path for different ribbons.

Table1 : Conductance measured on different GNRs. The absolute values, the actual probe spacing L as well as the temperature are given.

| Number | Conductance (µS) | Conductance ($G_0$) | Probe spacing (µm) | Temperature (K) |
|---|---|---|---|---|
| 1  | 58.479 | 1.509 | 0.50 | 298 |
| 2  | 36.101 | 0.931 | 5.00 | 298 |
| 3  | 34.965 | 0.902 | 5.00 | 298 |
| 4  | 49.751 | 1.284 | 1.50 | 298 |
| 5  | 41.152 | 1.062 | 5.00 | 298 |
| 6  | 45.871 | 1.184 | 1.50 | 298 |
| 7  | 29.673 | 0.766 | 5.00 | 298 |
| 8  | 34.722 | 0.896 | 5.00 | 298 |
| 9  | 36.900 | 0.953 | 0.50 | 298 |
| 10 | 40.012 | 1.032 | 1.50 | 298 |
| 11 | 48.309 | 1.247 | 5.00 | 298 |
| 12 | 50.021 | 1.291 | 1.50 | 298 |
| 13 | 37.878 | 0.978 | 5.00 | 298 |
| 14 | 38.610 | 0.997 | 5.00 | 298 |
| 15 | 43.290 | 1.117 | 1.50 | 298 |
| 16 | 44.444 | 1.147 | 5.00 | 298 |

| 17 | 44.843 | 1.158 | 0.50 | 298 |
| 18 | 48.780 | 1.259 | 5.00 | 298 |
| 19 | 55.865 | 1.442 | 0.50 | 298 |
| 20 | 62.111 | 1.603 | 0.50 | 298 |
| 21 | 37.593 | 0.970 | 1.50 | 120 |
| 22 | 35.460 | 0.915 | 5.00 | 120 |
| 23 | 37.453 | 0.967 | 5.00 | 120 |
| 24 | 41.667 | 1.076 | 1.50 | 120 |
| 25 | 43.290 | 1.117 | 5.00 | 120 |
| 26 | 42.735 | 1.103 | 1.50 | 120 |
| 27 | 46.082 | 1.190 | 1.50 | 120 |
| 28 | 56.179 | 1.450 | 0.50 | 120 |
| 29 | 59.171 | 1.527 | 0.50 | 120 |
| 30 | 54.644 | 1.411 | 1.50 | 120 |
| 31 | 39.920 | 1.030 | 5.00 | 120 |
| 32 | 39.904 | 1.030 | 1.50 | 120 |
| 33 | 38.910 | 1.004 | 5.00 | 120 |
| 34 | 45.248 | 1.168 | 0.50 | 120 |
| 35 | 40.322 | 1.041 | 1.50 | 120 |
| 36 | 40.512 | 1.046 | 1.50 | 32 |
| 37 | 35.971 | 0.929 | 5.00 | 32 |
| 38 | 31.152 | 0.804 | 5.00 | 32 |
| 39 | 35.842 | 0.925 | 5.00 | 32 |
| 40 | 37.735 | 0.974 | 5.00 | 32 |
| 41 | 37.879 | 0.978 | 1.50 | 32 |
| 42 | 40.160 | 1.037 | 5.00 | 32 |
| 43 | 38.167 | 0.985 | 5.00 | 32 |
| 44 | 44.052 | 1.137 | 0.50 | 28 |
| 45 | 43.459 | 1.122 | 0.50 | 28 |
| 46 | 35.842 | 0.925 | 5.00 | 298 |
| 47 | 40.485 | 1.045 | 5.00 | 298 |
| 48 | 39.797 | 1.027 | 1.50 | 78 |
| 49 | 39.370 | 1.016 | 1.50 | 78 |
| 50 | 37.693 | 0.973 | 0.50 | 78 |

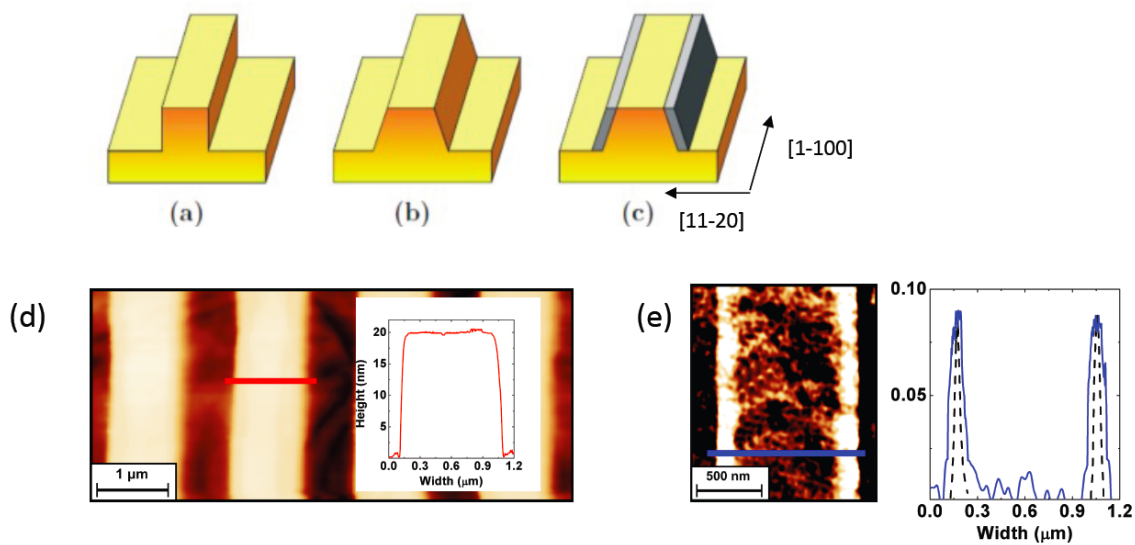

**Figure S10**
(a)-(c) Schematics of the MESA before (a) and after facet formation at 1420 K (b) and GNR formation at 1570 K(c). (d) AFM image of a MESA before annealing. The line scan demonstrates the successful fabrication of steep trench structures with well-defined etching depths of 20 nm. (e) EFM image taken after the final temperature step showing preferential growth of GNR at the step edges of the mesa. The dashed curve is obtained after de-convoluting of the AFM-tip shape.

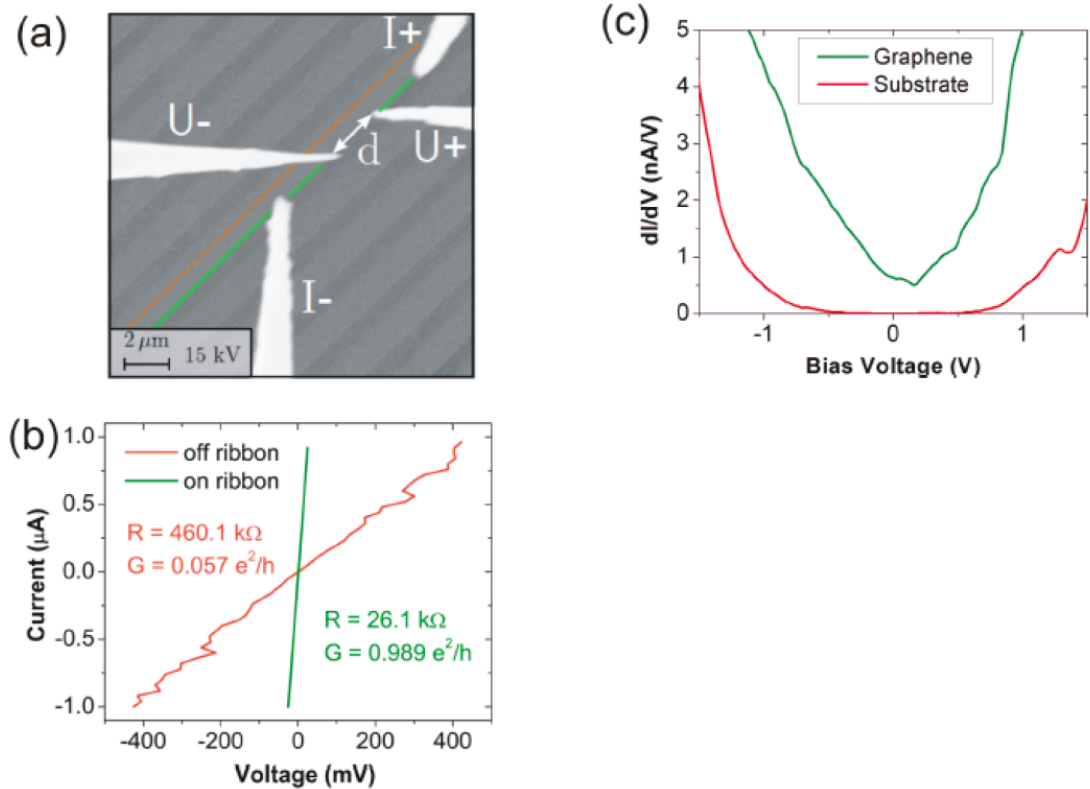

**Figure S11**
Demonstration of the collinear 4-probe in-situ resistance measurements. (a) SEM image of the tip positioning on top of a GNR. A current (typically +/-1µA) was passed to the nanostructure by using the outermost tips, while the voltage drop was measured with the inner two probes. The selective growth of GNR at the step edges is demonstrated with (b) 4-tip transport (1 µm tip distance), showing that the resistance on the terrace is 20 times higher than on the ribbon and (c) STS (set point 1 nA, 2 V, measured with lock-in technique). All measurements were done at room temperature. The color codes in the different graphs correspond to each other.

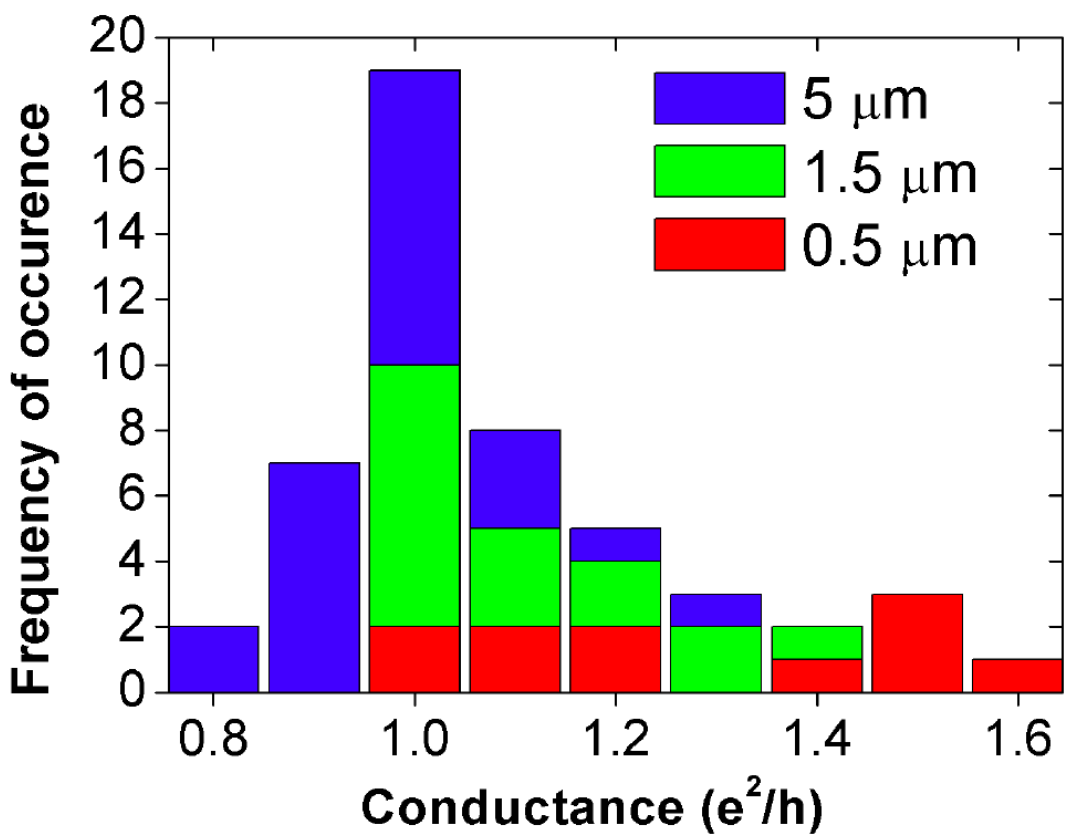

**Figure S12**
Histogram of the conductance values taken on 50 different GNRs and probe spacing (L=0.5μm, 1.5μm, 5μm) revealing clearly a peak at the conductance quantum $e^2/h$. There is a clear trend, that the smaller (larger) conductance values correspond to larger (smaller) probe spacing. The variation with temperature is extremely small (see table) and not shown here.

**References for supplementary information**

1. Stolyarova, E. et al. High-resolution scanning tunneling microscopy imaging of mesoscopic graphene sheets on an insulating surface. *Proceedings of the National Academy of Sciences of the United States of America* **104**, 9209-9212 (2007).
2. Miller, D.L. et al. Observing the Quantization of Zero Mass Carriers in Graphene. *Science* **324**, 924-927 (2009).
3. Knox, K.R. et al. Spectromicroscopy of single and multilayer graphene supported by a weakly interacting substrate. . *ArXiv: 0806.0355. Physical Review B* **78**, 201408(R) (2008).

4. Knox, K.R. et al. Making angle-resolved photoemission measurements on corrugated monolayer crystals: Suspended exfoliated single-crystal graphene. . *Physical Review B* **84**, 115401 (2011).
5. Sprinkle, M. et al. First Direct Observation of a Nearly Ideal Graphene Band Structure. *Physical Review Letters* **103**, 226803(4pp) (2009).
6. Hicks, J. et al. A wide-bandgap metal–semiconductor–metal nanostructure made entirely from graphene. *Nature Physics* **9**, 49-54 (2013).
7. Lin, Y.M., Perebeinos, V., Chen, Z.H. & Avouris, P. Electrical observation of subband formation in graphene nanoribbons. *Physical Review B* **78**, 161409(R) (2008).
8. Han, M.Y., Brant, J.C. & Kim, P. Electron Transport in Disordered Graphene Nanoribbons. *Physical Review Letters* **104**, 056801 (2010).
9. Wang, X.R. et al. Graphene nanoribbons with smooth edges behave as quantum wires. *Nature Nanotechnology* **6**, 563-567 (2011).
10. Todd, K., Chou, H.T., Amasha, S. & Goldhaber-Gordon, D. Quantum Dot Behavior in Graphene Nanoconstrictions. *Nano Letters* **9**, 416-421 (2009).
11. Schonenberger, C., Bachtold, A., Strunk, C., Salvetat, J.P. & Forro, L. Interference and Interaction in multi-wall carbon nanotubes. *Applied Physics a-Materials Science & Processing* **69**, 283-295 (1999).
12. Buttiker, M. Four terminal phase coherent conductance. *Physical Review Letters* **57**, 1761 (1986).
13. de Picciotto, R., Stormer, H.L., Pfeiffer, L.N., Baldwin, K.W. & West, K.W. Four-terminal resistance of a ballistic quantum wire. *Nature* **411**, 51-54 (2001).
14. Sprinkle, M. et al. Scalable templated growth of graphene nanoribbons on SiC. *Nature Nanotechnology* **5**, 727-731 (2010).
15. Norimatsu, W. & Kusunoki, M. Formation process of graphene on SiC (0001). *Physica E-Low-Dimensional Systems & Nanostructures* **42**, 691-694 (2010).
16. de Heer, W.A. et al. Large area and structured epitaxial graphene produced by confinement controlled sublimation of silicon carbide. *Proc Nat Acad Sci* **108**, 16900-16905 (2011).
17. Emtsev, K.V. et al. Initial stages of the graphite-SiC(0001) interface formation studied by photoelectron spectroscopy *Materials Science Forum* **55**, 6525 (2006).